\documentclass[prb, twocolumn, amsmath, amssymb]{revtex4-2}

\setcitestyle{numbers,square}
\usepackage{amsmath,amssymb}
\usepackage{tabularx}
\usepackage{graphicx}
\usepackage{bmpsize}
\usepackage{bm}
\usepackage{color}
\usepackage{amssymb}
\usepackage[version=3]{mhchem}
\usepackage{yhmath}
\usepackage{comment}

\DeclareMathAlphabet{\mathpzc}{OT1}{pzc}{m}{it}
\newcommand{\nn}{\nonumber}

\newcommand{\blue}[1]{\textcolor{black}{ #1}}
\newcommand{\mage}[1]{\textcolor{black}{ #1}}

\begin{document}
\makeatletter
\renewcommand\@biblabel[1]{[#1]}
\makeatother

\preprint{APS/123-QED}

\title{Singular flat bands in three dimensions: \\
Landau level spreading, quantum geometry, and Weyl reconstruction}

\author{Takuto Kawakami}
\email{kawakami@phys.sci.osaka-u.ac.jp}
\affiliation{Department of Physics, Osaka University, Toyonaka, Osaka 560-0043, Japan}
\author{Yuji Igarashi}
\affiliation{Department of Physics, Osaka University, Toyonaka, Osaka 560-0043, Japan}
\author{Mikito Koshino}
\affiliation{Department of Physics, Osaka University, Toyonaka, Osaka 560-0043, Japan}

\date{\today}

\begin{abstract}
We theoretically investigate three-dimensional singular flat band systems, 
focusing on their quantum geometric properties and response to external magnetic fields. 
As a representative example, we study the pyrochlore lattice, which hosts a pair of degenerate flat bands touching a dispersive band. 
We derive a three-orbital effective continuum model that captures the essential features near the band-touching point.
Within this framework, we identify the point-like topological singularity on a planar manifold defined by the degenerate flat band eigenvectors. 
This singularity strongly influences the quantum geometry and results in a characteristic Landau level structure, where the levels spread over a finite energy range.
We show that this structure reflects the underlying band reconstruction due to the orbital Zeeman effect,  
which lifts the flat band degeneracy and induces the Weyl-semimetal-like dispersion near the singularity. 
Our analysis reveals that the range of Landau level spreading is proportional to the quantum metric of each Zeeman-split band. 
We further demonstrate that adding a small dispersion via longer-range hopping induces unconventional Shubnikov-de Haas oscillations with a slowly drifting period. Finally, we show that our approach extends naturally to systems with higher orbital angular momentum, indicating the robustness of these features in a broad class of three-dimensional flat band models.
\end{abstract}


\maketitle

\section{Introduction}

Electronic systems featuring flat bands, where the bandwidth is vanishingly small compared to other relevant energy scales, have long been of interest in condensed matter physics, driven by their unusual single-particle properties~\cite{sutherland1986,bergman2008,guo2009, rhim2020,hwang2021geometric}, strongly correlated phases~\cite{tasaki1998review,lieb1989, mielke1991, mielke1991L73, mielke1992, tasaki1992,mielke1993,mielke1993example, bergholtz2013review,parameswaran2013review,volovik1994,yudin2014,volovik2018,tang2011,sun2011,neupert2011,zliu2013}, and experimental realizations~\cite{yin2022review, kim2017, lin2018, cao2018sc,cao2018,li2018, chen2019, yin2019, kang2020,ghimire2020review, kang2020ncom, wang2020,zliu2020, andrei2020,regan2020,tang2020,li2021, park2021,sun2022,yang2022,yang2023,di2023,ye2024}.
A flat band can arise trivially when interatomic hopping is suppressed, leading to a so-called trivial flat band. In contrast, nontrivial flat bands emerge despite finite hopping amplitudes, as a result of destructive interference encoded in the lattice geometry. Prototypical examples include the dice~\cite{sutherland1986}, Lieb~\cite{lieb1989}, and kagome~\cite{mielke1991L73} lattices. In such systems, the flat band is generally not isolated but instead touches a dispersive band at discrete points in momentum space. 
Considerable theoretical effort has been devoted to elucidating the mathematical origins and topological characteristics of these nontrivial flat bands~\cite{sutherland1986, mielke1991,mielke1991L73,mielke1992,tasaki1998review,nishino2003,nishino2005,bergman2008,hwang2021,hwang2021general,cualuguaru2022,hliu2022, kim2023, htli2025}.

Recent studies have highlighted the role of nontrivial topology and quantum geometry in flat-band systems~\cite{rhim2021review,bergman2008,rhim2019,rhim2020,oh2022,filusch2023,udagawa2024,song2025,penttila2025,yang2024,penttila2025,wyang2025}, particularly in two-dimensional (2D) settings.
\blue{Among these, singular flat bands~\cite{rhim2020,rhim2021review,hwang2021geometric,rhim2019} have attracted growing attention. 
These are flat bands that remain exactly flat while touching a dispersive band at a singular point that cannot be removed without breaking the flatness.
This singularity is protected by a discontinuity in the Bloch eigenvectors and is not merely a mathematical feature; it also has direct physical consequences.}
A striking effect appears in the Landau level structure under an external magnetic field. 
In a trivial flat band, the electronic states remain degenerate in the presence of a magnetic field, as the localized electrons are insensitive to the Aharonov–Bohm phase.
By contrast, in a singular flat band, the magnetic field induces Landau levels spreading over a finite energy range reflecting the underlying geometric singularity~\cite{rhim2020,rhim2021review}.

\begin{figure}[b]
\begin{center}
	\includegraphics[width=85mm]{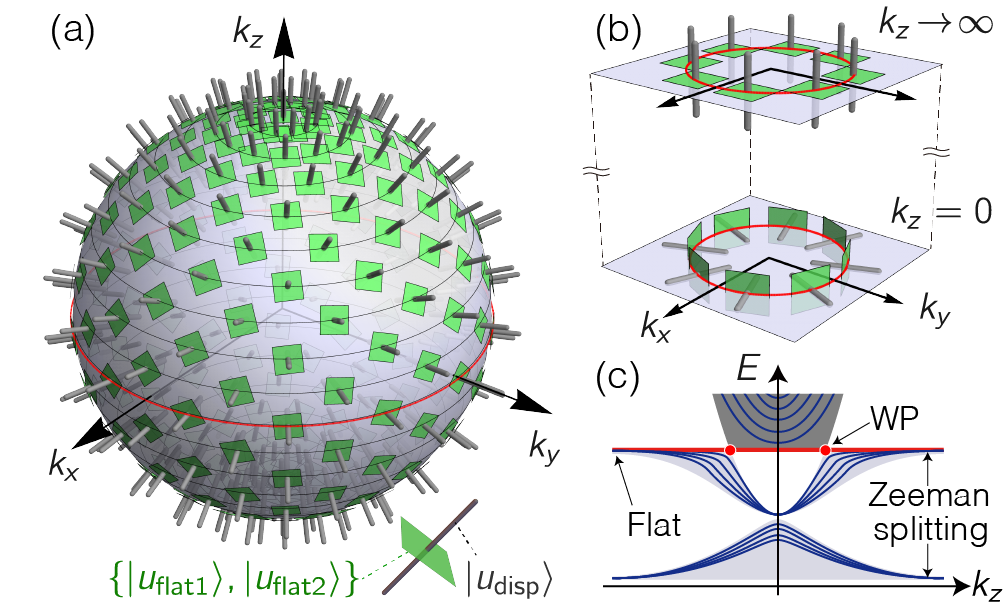}
	\caption{
	(a) Schematic diagram for point like singularity emerging in pyrochlore lattice. The green plane indicate the flat band manifold spanned by degenerate eigenvectors. The director perpendicular to the plane correspond to a eigenvector of the dispersive band. 
    (b) Asymptotic behavior of the flat band manifold for $k_z=0$ and $k_z\rightarrow\infty$. (c) Schematic Landau level structure, induced by orbital Zeeman effect that forms Weyl point (WP) in the continuum dispersion relation.
	}\label{fig:schematic}
\end{center}
\end{figure}

\blue{A natural question is whether singular flat bands can exist in three dimensions (3D) and, if so, how they respond to an external magnetic field. The existence of 3D singular flat bands was first demonstrated as an extension of 2D studies~\cite{rhim2019}. 
A notable candidate is the pyrochlore lattice model, which hosts flat bands extending along all three momentum directions  $(k_x,k_y,k_z)$, with a band touching point at $\bm{k}=0$~\cite{mielke1991L73, bergman2008,guo2009}. 
This structure can be regarded as a 3D analog of the kagome lattice.
Desipite its potential significance, the quantum geometric properties of these 3D flat bands remain largely unexplored.}

In this work, we investigate the geometric singularity of the 3D flat band in the pyrochlore lattice and examine its response to an external magnetic field.
Our analysis is based on a three-orbital effective continuum model, 
which captures the electronic structure near the threefold-degenerate band-touching point.
We show that the touching point is a point-like topological singularity in three-dimensional momentum space, as illustrated in Fig.~\ref{fig:schematic}(a). At any $k$-point, the eigenvectors can be represented by three-component real vectors. Notably, the flat band in the 3D system is twofold degenerate unlike in 2D cases, 
and hence its eigenvectors span a two-dimensional subspace within the three-component vector space, illustrated as green flat planes in Fig.~\ref{fig:schematic}(a).
The presence of the point-like singularity is manifested in a hedgehog structure of a normal director perpendicular to the plane of the flat-band manifold.
Additionally, the quantum geometric tensor for the flat bands acquires a non-Abelian structure~\cite{wilkzek1984,ma2010,palumbo2021}: each of its component is 
a $2\times 2$ matrix, reflecting the intrinsic degeneracy of the flat bands.

In the presence of external magnetic field applied along $z$-direction, 
the 3D system can be viewed as a collection of the effective 2D subsystems, 
labeled by $k_z$.
The quantum geometric properties of these 2D slices vary continuously with $k_z$ [see Fig.~\ref{fig:schematic}(b)]. 
At  $k_z = 0$, where the band-touching point is located, the normal director field winds around the origin, leading to a maximum in the quantum metric. In the limit of large $|k_z|$, the directors become parallel, resulting in a trivial geometry.

The three-dimensionality and intrinsic degeneracy introduce new features in the Landau level structure of the flat bands that are absent in 2D systems. 
For sufficiently large $|k_z|$, the degenerate flat bands are split into two branches by the orbital Zeeman effect, and the Landau levels within each branch remain nearly degenerate without energy spreading [Fig.~\ref{fig:schematic}(c)]. As $k_z$ decreases, the Landau levels in each branch acquire a finite energy spread, consistent with the increasing quantum metric. Near $k_z \sim 0$, hybridization between the two flat bands and the adjacent dispersive band gives rise to an intricate Landau level structure, potentially reflecting non-Abelian quantum geometry.
We show that the full $k_z$-dependent behavior is well captured by an effective band model that incorporates the orbital Zeeman effect while neglecting full Landau quantization. Within this approximation, the system effectively behaves as a Weyl semimetal, and the Landau level spreading can be interpreted as a dispersion of the flat bands induced by orbital magnetism. The model also explains the presence of a chiral Landau level connecting the flat bands to the dispersive band --- a characteristic feature of Weyl semimetals~\cite{nielsen1983,murakami2007, zyuzin2012, armitage2018weylreview, swang2017review, tch2016}.

Furthermore, when the flat band acquires a slight dispersion due to the further hopping term in the zero-field band structure, 
competition arise between the this weak dispersion and 
the spread-and-split structure associated with the Weyl semimetal states. 
We demonstrate that this interplay gives rise to a non-monotonic Landau level evolution as a function of $k_z$, resulting in unconventional Shubnikov–de Haas (SdH) oscillations with a slowly drifting oscillation period. 
This behavior can be attributed to a magnetic-field-dependent evolution of the Fermi surface, driven primarily by the orbital Zeeman effect.

Lastly, while our effective continuum model for the pyrochlore lattice is constructed from three orbitals correponding to orbital angular momentum $\ell=1$, 
the framework can be naturally extended to arbtirary angular momentum quantum numbers $\ell$, described by $2\ell+1$ dimensional matrix. 
In particular, $\ell=3/2$ corresponds to Luttinger-Kohn-type model with parameters tuned to yield a flat band. 
While the original Luttinger-Kohn model with $\ell=3/2$ describes dispersive bands and has been studied in the context of various topological materials~\cite{luttinger1956, luttinger1955,murakami2004, bernevig2006,dai2008,moon2013,savary2014,kondo2015,herbut2014,ruan2016, brydon2016, ruan2016prl, xu2017, bcheng2017,goswami2017,dzhang2018,venderbos2018,boettcher2018,ghorashi2018, kim2018, roy2019, kobayashi2019, sim2020, szabo2021, pzhu2024}, our work focuses on a flat band variant that can be generalized to arbitrary $\ell$. 
The extension demonstrates that the essential 
features of the 3D singular flat band identified in pyrochlore lattice, 
such as Landau level spreading and emergent Weyl semimetal like behevior, are robust across a wide class of band models.

The paper is organized as follows. 
In Sec.~\ref{sec:nonmag}, we discuss the 3D flat bands in field-free pyrochlore lattice and their singularity. 
In Sec.~\ref{sec:ll}, we formulate the Landau quantization of the effective Hamiltonian derived in Sec.~\ref{sec:nonmag} and clarifies the resulting characteristic level structure.
In Sec.~\ref{sec:sdh}, we analyze the Shubnikov-de Haas oscillation in the density of states as a possible signature of the obtained landau level structure. 
Finally, we demonstrate that our analysis can be systematically generalized into a wider class of the flat band system within the Luttinger-Kohn-type model with higher angular momentum in Sec.~\ref{sec:higherl}.

\section{Singular flat bands in Pyrochlore lattice}~\label{sec:nonmag}

\subsection{Effective continuum model}

\begin{figure}[b]
\begin{center}
	\includegraphics[width=85mm]{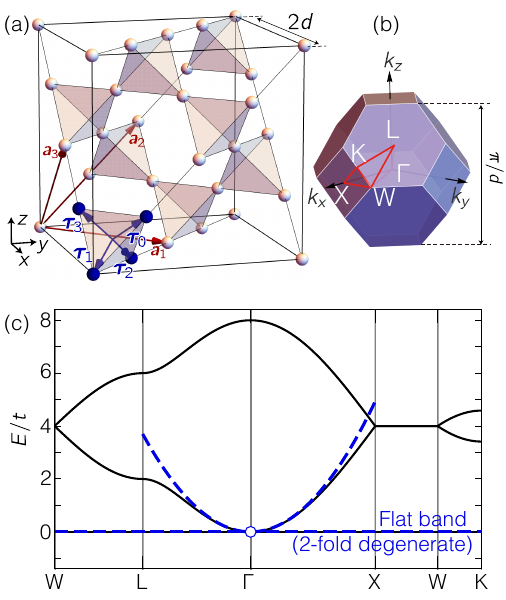}
	\caption{
	(a) Lattice structure of the pyrochlore lattice. Sublattices within a primitive unit cell is indicated in dark blue. 
	(b) Brillouin zone with high-symmetry point. 
	(c) Band structure of Pyrochlore lattice along high-symmetry lines in the Brillouin zone.
	The dashed curves in (c) is obtained from the effective continuum model Eq.~\eqref{eq:heffl}.
	}\label{fig:pyrochlore_lattice}
\end{center}
\end{figure}

We consider a pyrochlore lattice, as illustrated in Fig.~\ref{fig:pyrochlore_lattice}(a).
The primitive lattice vectors are defined as
$\bm{a}_1=d(\hat{\bm{x}}+\hat{\bm{y}})$, 
$\bm{a}_2=d(\hat{\bm{y}}+\hat{\bm{z}})$, and
$\bm{a}_3=d(\hat{\bm{z}}+\hat{\bm{x}})$, 
where $\hat{\bm{x}}$, $\hat{\bm{y}}$, and $\hat{\bm{z}}$ are unit vectors 
along $x$-, $y$-, and $z$-directions, respectively.
The nearest neighbor intersite distance is $d/\sqrt{2}$.
The unit cell consists of four sublattices with positions measured from the unit cell center [Fig.~\ref{fig:pyrochlore_lattice}(a)],
\begin{align}
\bm{\tau}_0 &= \tfrac{d}{4}(1,1,1), \quad &\bm{\tau}_1 &= \tfrac{d}{4}(1,-1,-1), \nonumber\\ 
\bm{\tau}_2 &= \tfrac{d}{4}(-1,1,-1), \quad &\bm{\tau}_3 &= \tfrac{d}{4}(-1,-1,1).
\end{align}

To describe electronic states in this system, 
we employ a nearest-neighbor tight-binding model,
\begin{equation}
H= t\sum_{\langle i,j\rangle} c^{\dag}_{i} c_{j}
\end{equation}
where $c_i$ indicates the annihilation operator for an electron at the 
lattice site $i$, 
and $\langle i,j\rangle$ represents a summation over all nearest neighboring atomic sites 
with hopping amplitude $t$. 
We define the Bloch basis as
\begin{equation}
|\bm{k},\bm{\tau}_X \rangle=\frac{1}{\sqrt{N_{\mathrm{c}}}}\sum_{\bm{R}_{X}} e^{i\bm{k}\cdot\bm{R}_{X}}|\bm{R}_X\rangle,
\end{equation}
where 
$\bm{R}_{X}=\bm{\tau}_X + l_1\bm{a}_1 + l_2 \bm{a}_2+ l_3 \bm{a}_3\, (l_i\in \mathbb{Z})$ 
represents atomic positions of $\bm{\tau}_{X}$-sublattice ($X=0,1,2,3$),
$N_{\mathrm{c}}$ is the number of the unit cell in the system.
The Bloch Hamiltonian takes the form,
\begin{align}\label{eq:bloch44}
(H_{\bm{k}})_{X,X'} &:= \langle \bm{k},\bm{\tau}_X|H | \bm{k},\bm{\tau}_{X'} \rangle \nn\\
	&= 2 t \cos \bm{k}\cdot{(\bm{\tau}_X-\bm{\tau}_{X'})},
\end{align}
where $X$, $X'$ are sublattice indices.
Here we added an on-site energy $V=2t$ to every site of the model, to shift the flat band to zero energy.
By diagonalizing Eq.~\eqref{eq:bloch44}, we obtain the band structure in Fig.~\ref{fig:pyrochlore_lattice}(c). 
We observe twofold degenerate three-dimensional flat band, which touches a dispersive band at the $\Gamma$ point.

We derive the low-energy effective continuum Hamiltonian which describes the flat and dispersive bands around the band touching point at $\Gamma$.
We introduce an orthonormal basis, 
\begin{align}\label{eq:4orb}
&|\varphi_x(\bm{k})\rangle=\frac{1}{2}(|\bm{k},\bm{\tau}_0 \rangle
+|\bm{k},\bm{\tau}_1\rangle-|\bm{k},\bm{\tau}_2\rangle-|\bm{k},\bm{\tau}_3\rangle)\nn\\
&|\varphi_y(\bm{k})\rangle=\frac{1}{2}(|\bm{k},\bm{\tau}_0 \rangle
-|\bm{k},\bm{\tau}_1\rangle+|\bm{k},\bm{\tau}_2\rangle-|\bm{k},\bm{\tau}_3\rangle)\nn\\
&|\varphi_z(\bm{k})\rangle=\frac{1}{2}(|\bm{k},\bm{\tau}_0 \rangle
-|\bm{k},\bm{\tau}_1\rangle-|\bm{k},\bm{\tau}_2\rangle+|\bm{k},\bm{\tau}_3\rangle)\nn\\
&|\varphi_w(\bm{k})\rangle=\frac{1}{2}(|\bm{k},\bm{\tau}_0 \rangle
+|\bm{k},\bm{\tau}_1\rangle+|\bm{k},\bm{\tau}_2\rangle+|\bm{k},\bm{\tau}_3\rangle).
\end{align}
At the $\Gamma$ point ($\bm{k} = 0$), these states are eigenstates of the Hamiltonian, with eigenenergy $E = 0$ for $\varphi_x$, $\varphi_y$, and $\varphi_z$, and $E = 8t$ for $\varphi_w$.
Around $\bm{k} \simeq 0$, the Hamiltonian can be expanded up to second order in $\bm{k}$ as
\begin{equation}\label{eq:heff}
\langle \varphi_{\alpha}| H |\varphi_{\beta}\rangle \\
\simeq \frac{td^2}{2}
\left(\begin{array}{ccc}
k_x^2 & k_x k_y & k_x k_z \\
k_y k_x & k_y^2 & k_y k_z \\
k_z k_x & k_z k_y & k_z^2
\end{array}\right):= H_{\mathrm{eff}},  
\end{equation}
where $\alpha,\beta\in\{x,y,z\}$. The eigenenergies are given by $E = 0, 0, (td^2/2)|\bm{k}|^2$.

The Hamiltonian $H_{\mathrm{eff}}$ has a 3D rotational symmetry:
\begin{equation} \label{eq:rotsym}
	\hat{R} H_{\mathrm{eff}}(\bm{k}) \hat{R}^{-1} = H_{\mathrm{eff}}(\hat{R} \bm{k}),
\end{equation}
where $\hat{R}$ is an arbitrary $3 \times 3$ rotation matrix.
The corresponding angular momentum operators are given by
\begin{gather}
L_x=
\left(\begin{array}{ccc}
0 & 0 & 0 \\
0 & 0 & -i \\
0 & i & 0 \\
\end{array}\right),\
L_y=
\left(\begin{array}{ccc}
0 & 0 & i \\
0 & 0 & 0 \\
-i & 0 & 0 \\
\end{array}\right),\nn\\
L_z=
\left(\begin{array}{ccc}
0 & -i & 0 \\
i & 0 & 0 \\
0 & 0 & 0 \\
\end{array}\right),
\end{gather}
which act as generators of the rotation operators in Eq.~\eqref{eq:rotsym}.
These $L_i$ 
satisfy the standard algebra $[L_\alpha,L_\beta]=i\epsilon_{\alpha\beta\gamma}L_\gamma$, and
represent effective orbital  angular momentum associated with the sublattice degrees of freedom.
Using them, the Hamiltonian in Eq.~\eqref{eq:heff} can be rewritten as
\begin{align}\label{eq:heffl}
H_{\mathrm{eff}} 
&=  \frac{t d^2}{2} \left(k^2 - (\bm{k}\cdot\bm{L})^2 \right),
\end{align}
which will be central to our analysis of Landau quantization in Sec.~\ref{sec:ll}.

\subsection{Singularity of 3D flat band}\label{sec:singularity}
We now demonstrate that band touching between the 3D flat band and dispersive band in pyrochlore lattice corresponds to a singularity of the wave function in $k$-space,
protected by a nontrivial topological invariant, as schematically depcited in Fig.~\ref{fig:schematic}(a). 
The key feature is the presence of $PT$ symmetry in the effective Hamiltonian Eq.~\eqref{eq:heff}, which satisfies
\begin{equation}
  H^\ast_{\mathrm{eff}}(\bm{k})=H_{\mathrm{eff}}(\bm{k})
\end{equation}
This constraint ensures that the eigenvector $\vec{\Psi}_\nu(\bm{k})$ is real and mutually orthogonal. 
In our model, the eigenstates form a three-component real vector, where $\nu=\pm 1$ labels the two flat bands and $\nu=0$ the dispersive band. 
As illustrated in Fig.~\ref{fig:schematic}(a), $\vec{\Psi}_{\pm1}(\bm{k})$ lie in the plane perpendicular to $\bm{k}$, while $\vec{\Psi}_0(\bm{k})$ is aligned along $\bm{k}$-direction.

Since the overall sign of an eigenvector does not affect its physical properties, 
each eigenstates is more appropriately described as a director, 
a vector defined up to an overall sign. 
Accordingly, the flat band subspace at each momentum point 
is characterized by a two-dimensional plane 
spanned by the two degenerate flat-band states. 
The vector $\vec{\Psi}_0$, which lies orthogonal to this plane, thus acts as a normal director to the flat-band subspace [see inset of Fig.~\ref{fig:schematic}(a)].

This structure implies that the manifold of the flat-band states form a coset space $S^2/\mathbb{Z}_2$, 
analogous to the order parameter space of nematic liquid crystals.
The mapping from a closed surface in momentum space (e.g., a sphere surrounding the band-touching point) 
to this manifold is topologically classified by the homotopy group $\pi_2(S^2/\mathbb{Z}_2)\simeq \pi_2(S^2)=\mathbb{Z}$, 
which counts how many times the director field $\vec{\Psi}_0$ 
wraps the two-sphere.

We define a monopole charge $Q$ associated with this mapping.  
Letting $\theta_{0}(\bm{k})$ and $\phi_{0}(\bm{k})$ be the polar and azimuthal angles parametrizing the director field $\vec{\Psi}_0$, the charge is given by
\begin{equation}
Q=\frac{1}{4\pi} \int_{S^2} d^2k \epsilon_{\alpha \beta} \sin\theta_{0} (\partial_{k_\alpha}\theta_{0})(\partial_{k_\beta}\phi_{0})
\end{equation}
where $\epsilon_{\alpha\beta}$ is the antisymmetric Levi-Civita symbol, and $k_{\alpha}$, $k_{\beta}$ span local coordinates on the surface $S^2$ in the $k$-space. 
In our effective Hamiltonian Eq.~\eqref{eq:heff}, 
we find $Q=1$ when the integration surface encloses the band touching point at $\bm{k}=0$ and  $Q=0$ otherwise. 
This behavior is consistent with the three dimensional texture in Fig.~\ref{fig:schematic}(a).
The existence of this quantized charge confirms that 
the degeneracy, or equivalently gap closing, between the dispersive and flat band at $\bm{k}=0$ is protected by a topological singularity in the wave function.

Furthermore, the nontrivial topology ensures that quantum distance between flat-band states remains finite even in the limit where the momentum-space surface enclosing the band touching point $\bm{k}=0$ shrinks to a point. 
That is, the quantum distance does not vanish as the momentum space separation tends to zero.
This signals a singular behavior in the flat-band wave function and supports its classification as a singular flat band in three dimensions.

\section{Landau levels of the singular 3D flat bands}\label{sec:ll}
\subsection{Landau level structure}
\label{sec:llsubsec}
We investigate the electronic structure of the 3D flat band system described by Eq.~\eqref{eq:heffl} under a uniform magnetic field $\bm{B}$ applied along the $z$-direction.
We define the vector potential $\bm{A}$ satisfying $\bm{B} = \bm{\nabla} \times \bm{A}$.
The Hamiltonian is then obtained by a standard process replacing the momentum components $k_\alpha$ in Eq.~\eqref{eq:heffl} with $\pi_\alpha/\hbar$ ($\alpha = x, y$), 
where $\bm{\pi} = -i\hbar \bm{\nabla} + e\bm{A}$
and the quadratic term is replaced with the symmetrized product $k_\alpha k_\beta\rightarrow (\pi_\alpha\pi_{\beta}+\pi_\beta\pi_{\alpha})/(2\hbar^2)$.
The momentum component along the magnetic field, $k_z$, remains a quantum number.
We define the ladder operators
$a = (\pi_x - i\pi_y)/\sqrt{2\hbar e B}$ and $a^\dagger = (\pi_x + i\pi_y)/\sqrt{2\hbar e B}$,
which satisfy the commutation relation $[a, a^\dagger] = 1$, noting that $[\pi_x, \pi_y] = i\hbar e B$.
As a result, the Hamiltonian in the magnetic field is written as
\begin{align}\label{eq:heffb}
	H_{B}(k_z) = \frac{\hbar\omega_B}{2} \Bigg[L_z^2\left(a^\dag a + \frac{1}{2}\right) + (1-L_z^2) (k_z l_B)^2 \nn\\
		-   \frac{L_-}{\sqrt{2}}(2L_z-1)   k_z l_B a^\dag-   \frac{L_+}{\sqrt{2}}(2L_z+1) k_z l_B a\nn\\
		- \frac{L_-^2}{2}a^{\dag2} -  \frac{L_+^2}{2}a^2 \Bigg],
\end{align}
where \begin{equation}
l_B=\sqrt{\frac{\hbar}{eB}},\quad  \hbar\omega_B=t\left(\frac{d}{l_B}\right)^2,
\end{equation}
and $L_{\pm}=L_x\pm iL_y$ are the ladder operators for the orbital angular momentum.

\begin{figure}[t]
\begin{center}
	\includegraphics[width=85mm]{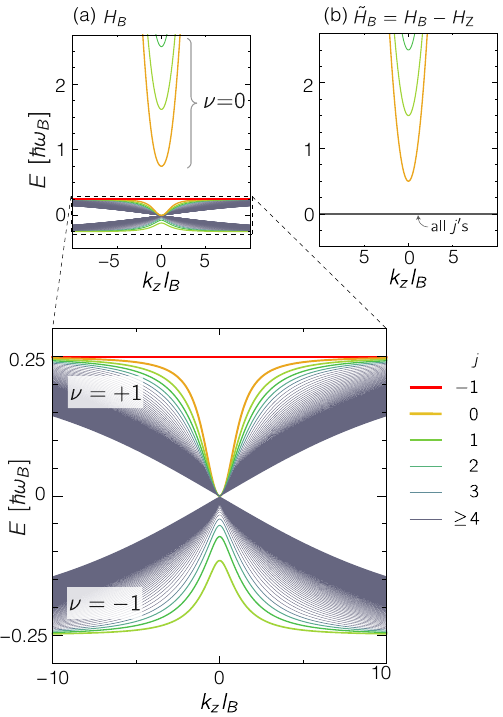}
	\caption{
	Landau level structure of the three dimensional singular flat band. 
	(a) Full spectrum as a function of $k_z$, including levels up to $j=j_{\mathrm{max}}$, with zoom in on the lower energy region to highlight feature associated with the flat band. We set $j_{\mathrm{max}}\sim l_B^2/d^2=100$.
    (b) Spectrum of the reduced Landau Hamiltonian $\tilde{H}_B=H_B-H_{\mathrm{Z}}$, where the orbital Zeeman term is subtracted. The black thick line indicates a set of flat levels, degenerate for all total angular momentum $j=-1,0,1,2,\cdots,$. 
	}\label{fig:llflat}
\end{center}
\end{figure}

The effective Hamiltonian in Eq.~\eqref{eq:heffb} satisfies the commutation relation
\begin{equation}\label{eq:comrel}
	[H_{\mathrm{B}}, J_z] = 0,
\end{equation}
where $J_z$ is the total angular momentum operator defined as
\begin{equation}
	J_z = a^\dag a + L_z.
    \label{eq_J_z}
\end{equation}
To describe the eigenstates of the Hamiltonian,  
we introduce basis states 
\begin{equation}\label{eq:basis}
|k_z, n, m\rangle
= \exp(ik_z z)\phi_{n}(x,y)\vec{e}_{m}.
\end{equation}
Here, $\phi_{n}(x,y)$ denotes the wave function of $n$-th Landau levels in two-dimensional system with $n= 0,1,2,\cdots$, where the index for the Landau level degeneracy in each $n$ is omitted.
The $\vec{e}_{m}$ is a three-dimensional eigenvector of the orbital angular momentum $L_z$ with the eigenvalue $m=-1,0,1$.
The basis of Eq.~\eqref{eq:basis} satisfies the following equations,
\begin{align}
    &a^\dag a |k_z, n, m\rangle = n |k_z, n, m\rangle, \nn \\
    &a|k_z,n,m\rangle 
=\sqrt{n}|k_z,n\!-\!1,m\rangle, \nn \\
&a^\dag|k_z,n,m\rangle 
=\sqrt{n+1}|k_z,n\!+\!1,m\rangle,
 \end{align}
 and 
 \begin{align}
&L_z |k_z, n, m\rangle = m|k_z, n, m\rangle, \nn\\
&L_{\pm}|k_z,n,m\rangle 
=\sqrt{(1\!\mp\! m)(2\!\pm\!m)}|k_z,n,m\!\pm\!1\rangle.
\end{align}

The commutation relation of Eq.~\eqref{eq_J_z} implies that the total angular momentum $j=n+m(\ge-1)$ is
conserved.
Accordingly, the eigenstates in the sector of $j$ and $k_z$ can be written as
\begin{equation}\label{eq:wfn}
	|\Psi_{\nu} \rangle = \sum_{m=-1}^{+1} \Psi_{\nu}(m)|k_z, j-m,m \rangle,
\end{equation}%
where $\nu$ is the index for the eigenstates in the sector. 
For $j\ge 1$, the eigenstate is expressed by a three-component vector 
$\vec{\Psi}_{\nu}=[\Psi_{\nu}(+1),\Psi_{\nu}(0),\Psi_{\nu}(-1)]^T$.
The corresponding Hamiltonian matrix is given as 
\begin{gather}
    H_{B}(j,k_z) = \tilde{H}_{B}(j,k_z) + H_\mathrm{Z} \label{eq:hj}.
\end{gather} 
where 
\begin{gather}
	\tilde{H}_{B} = 
	\frac{\hbar\omega_B}{2} \left(\begin{array}{ccc}
	j  & -\sqrt{j} k_z l_B   & - \sqrt{j (j+1)} \\
	-\sqrt{ j} k_zl_B  &  ( k_zl_B)^2  &  \sqrt{j+1}  k_zl_B  \\
	- \sqrt{j(j+1)} & \sqrt{j+1}  k_z l_B  &  j + 1
	\end{array}\right), \label{eq:hjflat}\\
	H_{\mathrm{Z}} = -\frac{\hbar \omega_B}{4} L_z.  \label{eq:hz}
\end{gather}
We separate $\bar{H}_B$ and $H_{\mathrm{Z}}$ for convenience in the subsequent analysis.
Note that $L_z$ in Eq.~\eqref{eq:hz} is expressed as 
$L_z = {\rm diag}(1,0,-1)$ in this basis.
The eigenenergies of the sector can be obtained by diagonalizing the $3\times 3$ matrix, $H_B$.

For $j = -1$ and $0$, the eigenvectors reduce to 
one- and two-component forms, respectively,  
since the basis states $|k_z, n, m\rangle$ with $n < 0$ vanish.  
Correspondingly, the Hamiltonian matrices for $j = -1$ and $j = 0$ are given by  
the $1 \times 1$ and $2 \times 2$ lower-right submatrices of Eq.~\eqref{eq:hj}, respectively.
The eigenenergies $E_\nu(j,k_z)$ of the $j=-1$ and $0$ sectors are
obtained analytically as
\begin{gather}
	E_{+1}(-1,k_z) = \frac{\hbar \omega_B}{4}, \label{eq:ejm1} \\
	E_{\nu}(0,k_z) = \frac{\hbar \omega_B}{4} \Bigg(k_z^2l_B^2\!+\!\frac{3}{2}\!+\!(1\!-2\nu)\sqrt{\left(k_z^2l_B^2+\tfrac{1}{2}\right)^2+2}\Bigg)\nn\\ 
	  \qquad  \qquad \qquad \qquad \qquad \qquad \qquad \qquad(\nu = 1,0) \label{eq:ej0}
\end{gather}

The full Landau level spectrum (including all $j$) as a function of $k_z$ is shown in Fig.~\ref{fig:llflat}(a),  
where the lower panel presents a magnified plot of the flat-band region.
The resulting spectrum can be classified into three main branches, labeled by $\nu=-1,0,1$, where $\nu=\pm1$ correspond to flat bands, 
while $\nu=0$ corresponds to a dispersive branch.
We find that the flat bands spread over finite energy range, $-\hbar\omega_B/4<E<\hbar\omega_B/4$.
In the limit $k_z\rightarrow\infty$, 
all the $\nu=\pm 1$ levels converge to $\pm\hbar\omega_B/4$.
We note that the maximum total angular momentum is estimated as $j_{\mathrm{max}}\sim l_B^2/d^2$ ensuring conservation of the total state density (see Appendix~\ref{sec:dosderive}). 
In Fig.~\ref{fig:llflat}, we set $j_{\mathrm{max}}=100$, 
which corresponds to $B\sim12$ T for a typical pyrochlore material with $d\sim 5 $ \AA~\cite{hase2018}.
As $B$ decreases, $j_\mathrm{max}$ increases, and the gap edge of $\nu=+1$ ($\nu=-1$) branch shifts downward (upward) toward zero.

Notably, 
in the Landau level Hamiltonian with the Zeeman term  subtracted  i.e., $\tilde{H}_{B}(j,k_z)=H_{\mathrm{B}}(j,k_z)-H_{\mathrm{Z}}$,
the LL spreading of the flat band is completely suppressed recovering fully flat band, as shown in Fig.~\ref{fig:llflat}(b).
Here the energy levels are explicitly written as
\begin{gather}\label{eq:zeroth}
    \tilde{E}_{\pm 1}(j,k_z) = 0,\quad  \tilde{E}_{0}(j,k_z) = \hbar\omega_B \left(j+\frac{1}{2}+\frac{k_z^2l_B^2}{2}\right),
\end{gather}
where $\tilde{E}_{\pm 1}(j,k_z)$ are degenerate for all $k_z$ and $j$.
$\tilde{E}_{0}(j,k_z)$ corresponds to standard Landau level 
for a three dimensional free electron with zero-field eigenstates $E(\bm{k})=(td^2/2)|\bm{k}|^2$. 
This clearly indicates that the spreading of the Landau levels entirely originates from the orbital Zeeman term $H_{\mathrm{Z}}$.

Landau level spreading in large $k_z$  can be understood by a perturbation theory with respect to  $H_{\mathrm{Z}}$.
At the first order in $H_\mathrm{Z}$, the degeneracy between the flat bands with $L_z=\pm 1$ is lifted, shifting their energy to $\mp\hbar\omega_B/4$.
This explains the splitting of the flat-band Landau levels into two branches in the $k_z \to \infty$ limit, as seen in Fig.~\ref{fig:llflat}.
In this regime, the $\nu=\pm1$ and $0$ branches in Fig.~\ref{fig:llflat} correspond to angular momentum $L_z=\mp1$ and $0$, respectively.
As $k_z$ decreases, each branch of the flat-band Landau levels spreads.
Using the second-order perturbation theory, 
the asymptotic behavior in the limit $j\ll k_z l_B$ can be written as
\begin{equation}\label{eq:elargek}
\begin{split}
	E_{\pm1}(j,k_z) \approx \pm \frac{\hbar\omega_B}{4} \left( 1- \frac{j+1/2\pm1/2}{k_z^2l_B^2} \right) + \mathcal{O}\left(\frac{j}{k_z^4l_B^4}\right), \\
\end{split}
\end{equation}
This spreading behavior can also be described by the quantum metric, as will be discussed in Sec~\ref{sec:qg}.

As approaching $k_z = 0$, $\nu=+1$ levels rapidly contract to $E=0$, 
while the spread of $\nu=-1$ remains.
At $k_z=0$, the Hamiltonian Eq.~\eqref{eq:hj} is separated to
block-diagonal matrices for the $L_z=\pm 1$ and $L_z=0$. 
The $L_z=\pm 1$ sector is equivalent to the continuum model for the kagome lattice hosting a 2D singular flat band and a dispersive band \cite{rhim2020,rhim2021review}, and
the spread of the $\nu=-1$ Landau levels
is equivalent to the LL spreading of the 2D singular flat band \cite{rhim2020,rhim2021review}.
The $L_z=0$ sector corresponds to a 2D trivial flat band with no singularity, consistent with the fully degenerate levels in $\nu=+1$ branch.
As $k_z$ is continuously varied from $k_z = \infty$ to $0$, the flat bands  transition between different $L_z$ sectors: they are associated with $L_z = \pm 1$ in the $k_z \to \infty$ limit, whereas at $k_z = 0$, they involve $L_z = \pm 1$ and $0$.

\begin{figure}[t]
\begin{center}
	\includegraphics[width=85mm]{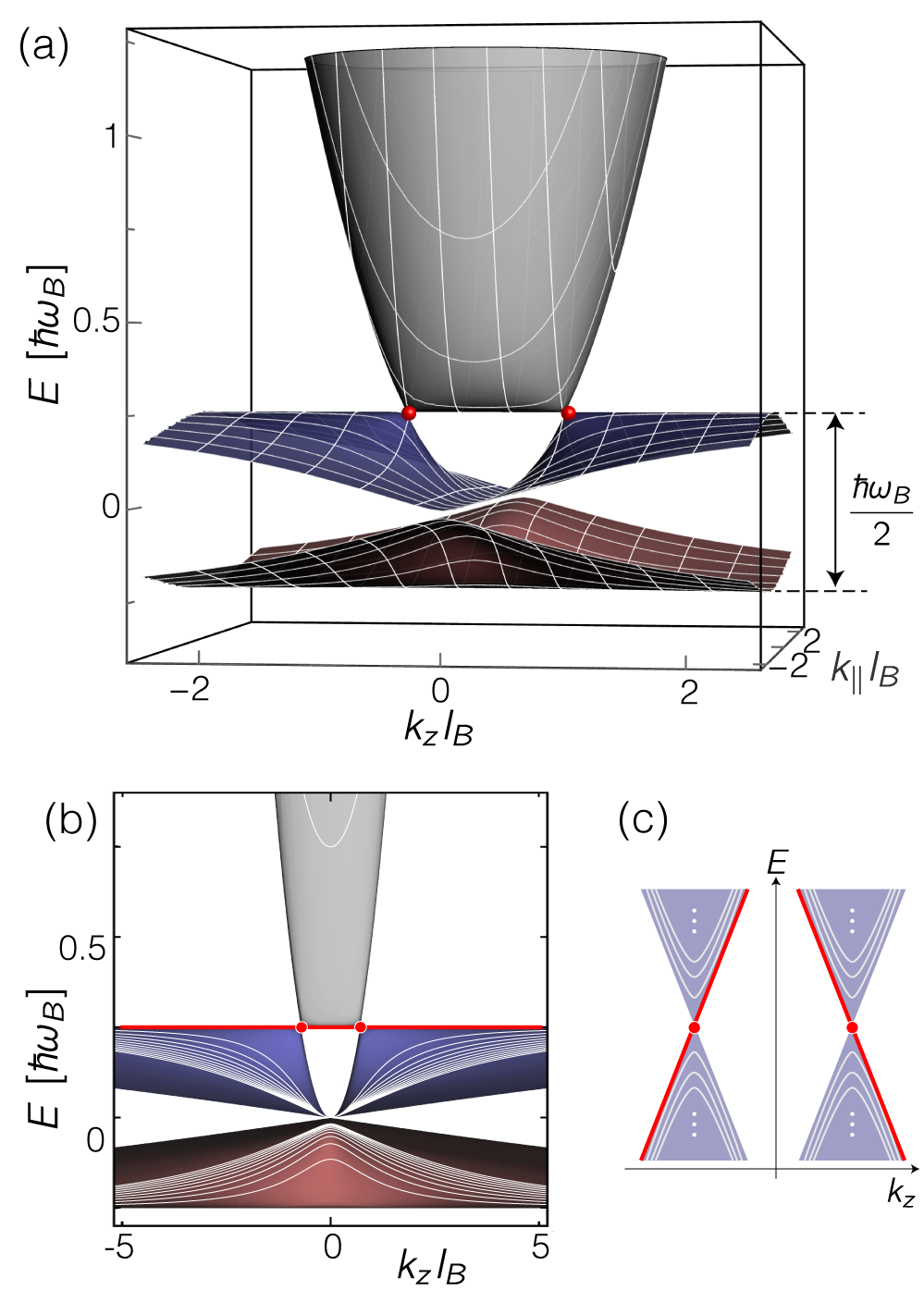}
	\caption{
	Weyl semimetallic band structure associated with the orbital Zeeman effect.
	(a) Energy spectrum of $H_{\mathrm{eff+Z}}=H_{\mathrm{eff}}+H_{\mathrm{Z}}$.
	(b) Landau levels for $-1\le j\le 10$, overlaid on the spectrum in (a). 
    (c) Schematic Landau level structure of a typical Weyl semimetal with the finite group velocity. 
	}\label{fig:weyl}
\end{center}
\end{figure}

\subsection{Effective Weyl semimetal model}
\label{sec:weyl}

\mage{
In conventional electronic systems, the Landau level spectrum can be approximately understood via semiclassical Bohr–Sommerfeld quantization, in which Landau levels emerge from the discretization of constant-energy contours of the original energy band.
In singular flat-band systems, however, this semiclassical framework does not straightforwardly account for the observed Landau level spreading,
as a dispersionless flat band leads to degenerate Landau levels at a single energy within this scheme.
In this section, we show that a band model incorporating only the orbital Zeeman effect provides an effective underlying continuous band dispersion that accounts for the spread of the Landau levels.
}

Specifically, we consider a modified Hamiltonian 
\begin{equation}\label{eq:hcontplusz}
 H_{\mathrm{eff+Z}} = H_{\mathrm{eff}} + H_{\mathrm{Z}}, 
\end{equation}
in which the orbital Zeeman term $H_{\mathrm{Z}}$ in Eq.~\eqref{eq:hz} is added to the continuum Hamiltonian $H_{\mathrm{eff}}$ in Eq.~\eqref{eq:heffl}, without replacing the momentum $\bm{k}$ with $\bm{\pi}/\hbar$.
Figure~\ref{fig:weyl}(a) shows the band dispersion of $H_{\mathrm{eff+Z}}$ as a function of $k_z$ and $k_\parallel$, where $k_\parallel$ denotes the in-plane momentum along an arbitrary direction in the $k_xk_y$ plane.
The resulting band structure reproduces the overall spectral features of the Landau levels as a function of $k_z$ shown in Fig.~\ref{fig:llflat}, except for the discrete level structure arising from the quantization of the $k_xk_y$ plane.
 In particular, the 3D flat band splits into $\pm \hbar \omega_B / 4$ in the regime $k_z l_B \gg 1$, and the upper flat-band branch becomes dispersionless at $k_z = 0$.

Most notably, the band structure in Fig.~\ref{fig:weyl}(a) exhibits a characteristic feature of a Weyl semimetal~\cite{nielsen1983,murakami2007, zyuzin2012, armitage2018weylreview, swang2017review, tch2016}, where the dispersive band and the upper flat band touch at two points in 3D momentum space,
$\bm{k}_{\pm}=\pm \hat{\bm{z}}/(\sqrt{2} l_B)$, where $\hat{\bm{z}}$ is a unit vector along $z$.
In general, the energy bands of Weyl semimetals comprise a pair of band-touching points with opposite helicities. In the presence of a magnetic field, a chiral zeroth Landau level emerges, connecting the valence and conduction bands as illustrated in Fig.~\ref{fig:weyl}(c).
In the present system, the chiral Landau level corresponds to the 0th level in the $j = -1$ sector, shown as the red flat line in Fig.~\ref{fig:llflat}, and schematically illustrated in Fig.~\ref{fig:weyl}(b) as a special case of the configuration in Fig.~\ref{fig:weyl}(c).

\begin{figure}[t]
\begin{center}
	\includegraphics[width=85mm]{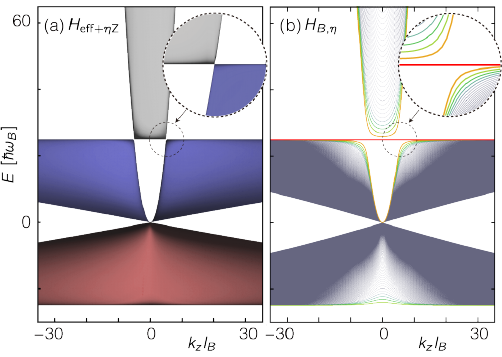}
	\caption{\blue{Comparison between the energy spectrum of the effective Weyl model and original Landau level structure, calculated with an enhanced Zeeman term. The enhancement factor is taken to be $\eta=50$.} 
	}\label{fig:enhance}
\end{center}
\end{figure}

\mage{
To better illustrate this correspondence, we introduce a modified model in which the Zeeman term is artificially enhanced. Specifically, we consider the continuum and quantized Landau level Hamiltonians defined as  
\begin{eqnarray}\label{eq:effetaz}
&H_{\mathrm{eff}+\eta\mathrm{Z}} = H_{\mathrm{eff}}(\bm{k}) + \eta H_{\mathrm{Z}},
\\
\label{eq:beta}
&H_{B,\eta}(j,k_z) = \tilde{H}_B + \eta H_{\mathrm{Z}},
\end{eqnarray}
where \(\eta\) is an artificial enhancement factor applied to the orbital Zeeman term.
The calculated band structures for \(\eta = 50\), obtained from Eqs.~\eqref{eq:effetaz} and \eqref{eq:beta}, are shown in Figs.~\ref{fig:enhance}(a) and \ref{fig:enhance}(b), respectively. The enhancement factor \(\eta\) effectively rescales the bandwidth of the originally flat-band cluster, while the Landau level spacing in the quantized model [Fig.~\ref{fig:enhance}(b)] remains essentially unchanged.
In Fig.~\ref{fig:enhance}(b), we clearly observe that the chiral Landau level (red curve) penetrates the two Weyl points of the continuum spectrum shown in Fig.~\ref{fig:enhance}(a). As \(\eta\) is reduced toward \(\eta = 1\), the spectrum continuously evolves back to that of Fig.~\ref{fig:weyl}(b), while retaining its key spectral features.
This analysis supports our conclusion that the emergence of the special flat Landau level in Fig.~\ref{fig:llflat} is a direct consequence of the Weyl semimetal structure induced by the orbital Zeeman effect.
}

\subsection{Quantum geometry and Landau level spreading}\label{sec:qg}

In two-dimensional singular flat-band systems such as the kagome lattice, the Landau level energy spread is known to be quantitatively linked to the underlying quantum geometry of the energy band~\cite{rhim2019,rhim2020,hwang2021geometric}. 
This raises the intriguing question of whether a similar geometric framework can be extended to 3D flat band systems.
In our 3D case, with a magnetic field applied along the $z$-direction, the system can be regarded as a collection of two-dimensional subsystems indexed by $k_z$.
Unlike the 2D counterparts~\cite{rhim2019,rhim2020,hwang2021}, however, the flat bands in this system remain doubly degenerate throughout the Brillouin zone, where the non-Abelian nature of the quantum geometric tensor~\cite{wilkzek1984,ma2010,palumbo2021} complicates the direct application of existing Abelian-based approaches.

This difficulty can be avoided by employing the effective Weyl semimetal model introduced in Eq.~\eqref{eq:hcontplusz}. In this framework, the degeneracy of the flat band is lifted by the Zeeman splitting, allowing the quantum geometric tensor to be evaluated for each branch individually. As demonstrated below, this procedure reveals that the Landau level bandwidth in the large-$k_z$ regime is proportional to the quantum metric.

We consider the effective model with the Zeeman splitting, $H_{\mathrm{eff+Z}}$ [Eq.~\eqref{eq:hcontplusz}]
in the regime $k_z\gg k_{\parallel}=\sqrt{k_x^2+k_y^2}$.
Expanding the Hamiltonian up to first order in $k_\parallel$, we obtain 
\begin{equation}
H_{\mathrm{eff+Z}} \approx H_{\mathrm{eff+Z}}^{(0)} + H_{\mathrm{eff+Z}}^{(1)}, 
\end{equation}
with
\begin{gather}
H_{\mathrm{eff+Z}}^{(0)}= \frac{td^2}{2}\left(k_z^2(1-L_z^2)-\frac{1}{2l_B^2}L_z\right) \nn\\
H_{\mathrm{eff+Z}}^{(1)}= \frac{td^2}{2}k_{\parallel}k_{z}
\Big(\cos\phi_k \{L_z,L_x\} +\sin\phi_k \{L_z,L_y\}\Big). 
\end{gather}
where $(k_x, k_y)=k_{\parallel}(\cos\phi_k,\sin\phi_k)$, 
and $\{X,Y\}=XY+YX$.

In the limit of $k_z \rightarrow \infty$, the Hamiltonian is dominated by the zeroth-order term $H_{\mathrm{eff+Z}}^{(0)}$, which depends solely on the orbital angular momentum operator $L_z$. As a result, the eigenstates of the Hamiltonian are simply the eigenstates of $L_z$, expressed as $|\psi^{(0)}_{m}\rangle = \vec{e}_{m}$, where $m = -1, 0, 1$ denotes the eigenvalue of $L_z$. In this regime, the eigenstates become independent of $k_x$ and $k_y$, indicating a flat geometry in the $k_x$–$k_y$ plane. Correspondingly, the director shown in Fig.\ref{fig:schematic}(b) aligns uniformly as $k_z \rightarrow \infty$. This trivial quantum metric reflects the vanishing Landau level spreading in the large-$k_z$ limit, as seen in Fig.\ref{fig:llflat}.

At finite $k_z$, the wave functions for the Zeeman-split flat bands ($m = \pm 1$) 
are perturbed by $H_{\mathrm{eff+Z}}^{(1)}$, and are given to first order in $k_\parallel$ as
\begin{equation}
|\psi_{\pm1}\rangle = \frac{1}{\sqrt{N_{\pm1}}}\left(\exp(\mp i\phi_k)\vec{e}_{\pm1} \mp \frac{1}{\sqrt{2}}\frac{k_\parallel k_zl_{B}^2 }{k_z^2l_{B}^2\pm 1/2} \vec{e}_0\right),
\end{equation}
where $N_{\pm 1}=1+k_{\parallel}^2k_z^2l_B^4/[2(k_z^2l_B^2\pm 1/2)^2]$. 
From these expressions, the quantum metric component along the azimuthal direction can be computed as
\begin{align} \label{eq:g}
g_{\phi_k\phi_k,\pm} =& \mathrm{Re}\big[ \langle\partial_{\phi_k} \psi_{\pm1}|\partial_{\phi_k} \psi_{\pm1}\rangle \nn\\
& -\langle\partial_{\phi_k} \psi_{\pm1}|\psi_{\pm1}\rangle\langle \psi_{\pm1}|\partial_{\phi_k} \psi_{\pm1}\rangle \big] \nn \\
\simeq &\frac{1}{\sqrt{2}} \left(k_{\parallel} / k_z\right)^2
\end{align}
to leading order in $k_{\parallel}/k_z$. 
\blue{Notably, this component is directly proportional to the maximum Hilbert Schmidt quantum distance between states on a circle at fixed $k_z$ and $k_\parallel$, which is given by
\begin{align}\label{eq:dmax}
    d_{\mathrm{max},\pm}&=\max_{\phi_{k},\phi_{k'}} (1-|\langle\psi_{\pm1}(\phi_{k'})|\psi_{\pm}(\phi_k)\rangle|) \nn\\ &\simeq (k_\parallel/k_z)^2 = \sqrt{2}{g_{\phi_k\phi_k,\pm}}
\end{align}}
In the large $k_z$ regime, the Landau level spreading is given by the second term of Eq.~\eqref{eq:elargek}, and it is proprtional to $j/k_z^2$ with the total angular momentum index $j$.
The metric component of Eq.~\eqref{eq:g}, \blue{as well as the maximum quantum distance in Eq.~\eqref{eq:dmax}}, is proportional to this band width, noting that
$j\sim n\sim k_{\parallel}^2$
for large $n\gg1$. 
This establishes a clear link between the quantum metric properties of the Zeeman-split flat band and calculated Landau level spreading.
At $k_z \lesssim k_\parallel$, the degenerate flat bands are strongly hybridized. 


\section{Shubnikov-de Haas Oscillations in bent flat bands}~\label{sec:sdh}


In the idealized nearest-neighbor tight-binding model of the pyrochlore lattice, the flat bands remain perfectly dispersionless. 
In real materials, however, longer-range hopping induces a slight bending of these bands, resulting in a finite bandwidth.
This intrinsic dispersion can obscure the Landau level spreading. Nevertheless, the topologically nontrivial nature of the singular flat band can still be probed through the observation of Shubnikov–de Haas (SdH) oscillations, as demonstrated below.

To account for this effect, we include an additional hopping term between sites at $\bm{\tau}_X$ and $\bm{\tau_}X \pm \bm{L}_i$ ($i = 1, \cdots, 6$), which belong to the same sublattice. Here, the relative vectors are defined as $\bm{L}_i = \bm{a}_i$ and $\bm{L}_{3+i} = \bm{a}_{i+1} - \bm{a}_{i+2}$ ($i = 1,2,3$), with the cyclic condition $\bm{a}_{i+3} = \bm{a}_i$. These additional hopping introduces an additional term to the Bloch Hamiltonian [Eq.~\eqref{eq:bloch44}], given by
\begin{equation}\label{eq:hprime}
(H_{\bm{k}}')_{X,X'} = 2t' \sum_{i=1}^{6} \big(-1 + \cos \bm{k} \cdot \bm{L}_i\big) \delta_{X,X'},
\end{equation}
where we have shifted the energy zero point by $-12t'$.
Accordingly, the continuum zero-field Hamiltonian [Eq.~\eqref{eq:heff}] 
and the Landau-level Hamiltonian [Eq.~\eqref{eq:heffb}] acquire correction terms,
\begin{gather}
	H'_{\mathrm{eff}} = -4 t' d^2 \bm{k}^2\\
	H'_{B} = - \frac{4t'}{t}\hbar\omega_B  \left(a^\dag a + \frac{1}{2}  + k_z^2l_B^2 \right)
\end{gather}
respectively.

\begin{figure}[t]
\begin{center}
	\includegraphics[width=85mm]{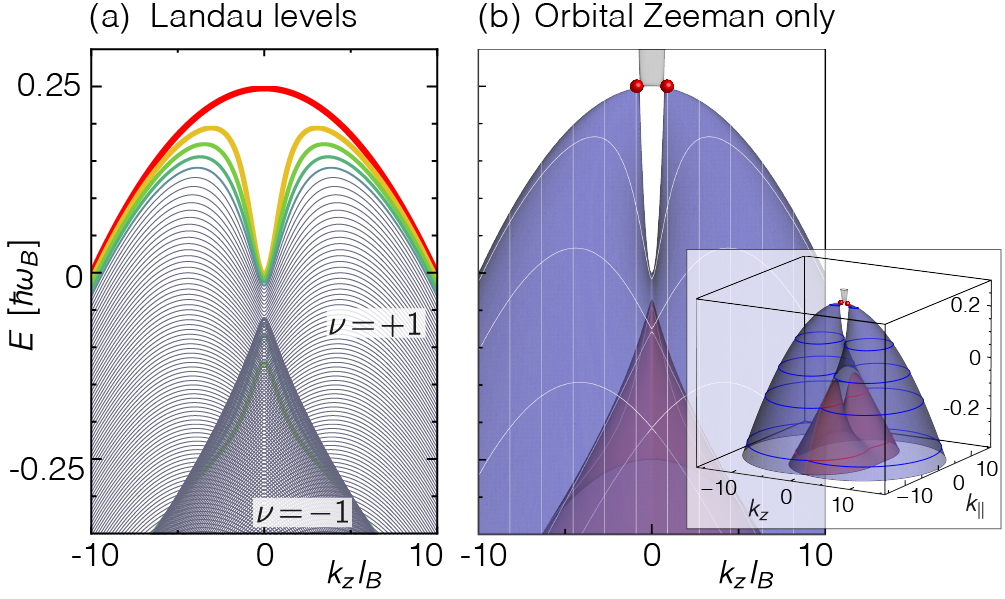}
	\caption{
	Electronic structure of slightly bended flat band under the magnetic field. 
        (a) Landau level structure. 
        (b) Dispersion relation of $H_{\mathrm{eff}} + H_{\mathrm{eff}}'+H_{Z}$, with Weyl points indicated by red spheres. 
        The inset shows a three-dimensional plot of (b) as a function of $k_z$ and $k_{\parallel}$. 
        The parameter $4t'/t=0.0025$ is used.
	}\label{fig:lldisp}
\end{center}
\end{figure}

Figure~\ref{fig:lldisp}(a) shows the Landau levels in the flat band region, calculated from $H_{\mathrm{B}} + H_{\mathrm{B}}'$ with $4t' = 0.0025 t$.  
For $t' > 0$, the Landau levels bend downward with increasing $k_z$.  
In Fig.~\ref{fig:lldisp}(b), we show the corresponding energy band of the effective model with the orbital Zeeman term,  
$H_{\mathrm{eff+Z}} = H_{\mathrm{eff}} + H_{\mathrm{eff}}' + H_{Z}$, where the Weyl points are indicated by red spheres.  
We observe good agreement between the Landau level structure and the Weyl semimetal band even in the presence of the additional band dispersion.
Here we note that the momentum $k_z$ (horizontal axis) and the energy $E$ (vertical axis) are scaled by $l_B$ and $\hbar\omega_B$, respectively, so that the plots in Figs.~\ref{fig:lldisp}(a) and (b) are independent of the magnetic field amplitude under this scaling.

\begin{figure}[t]
\begin{center}
	\includegraphics[width=85mm]{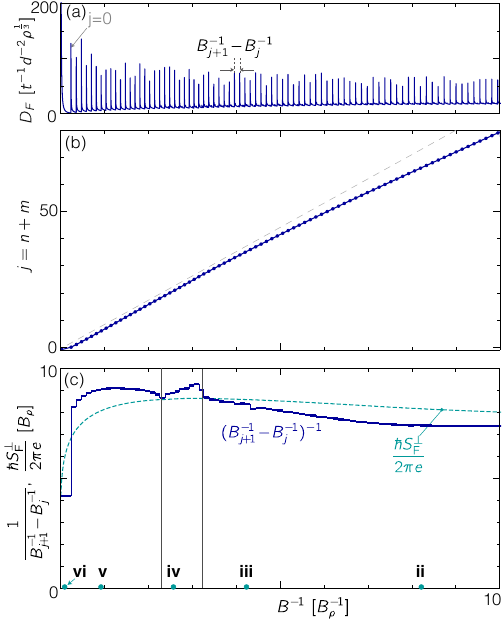}
	\caption{
	Shubnikov-de Haas oscillation and fermi surface evolution in pyrochlore lattice. (a) Oscillation in the density of states as a function of inverse magnetic field $B^{-1}$, obtained from the landau levels in Fig.~\ref{fig:lldisp}(a). (b) Angular momentum $j=n+m$ characterizing each peak in panel (a). Dashed line highlights the eventual change in slope. (c) Inverse peak spacing from panel a (a) (blue polyline), overlayed with the maximal cross section of the fermi surface $S_{F}^{\perp}$ (green dashed curve). 
	}\label{fig:oscillation}
\end{center}
\end{figure}

We examine the SdH oscillation when the Fermi energy lies slightly below the top of the bent flat band. We define $\rho\, (<0)$ as the charge density relative to the fully filled flat bands. The density of states at the Fermi level, $D_{\mathrm{F}}$, is computed for a fixed $\rho$, following the formulation in Appendix~\ref{sec:dosderive}. The resulting $D_{\mathrm{F}}$ as a function of inverse magnetic field $B^{-1}$ is shown in Fig.~\ref{fig:oscillation}(a), where the magnetic field is scaled in units of $B_{\rho} = (\hbar/e)|\rho|^{2/3}$. The plot exhibits a series of peaks, indicative of SdH oscillations in the conductivity.
The origin of these peaks can be understood in the standard manner.
As the magnetic field $B$ increases with fixed $\rho$, the Fermi energy $E_{\mathrm{F}}$ moves upward in the scaled Landau level spectrum shown in Fig.~\ref{fig:lldisp}(a), reflecting the increasing Landau level degeneracy. When $E_{\mathrm{F}}$ aligns with an extremum of a Landau level dispersion, the density of states diverges, giving rise to a sharp peak. Consequently, the position of the $j$-th peak, $B = B_j$, can be labeled by the total angular momentum index $j = n + m$.

\begin{figure*}[t]
\begin{center}
	\includegraphics[width=180mm]{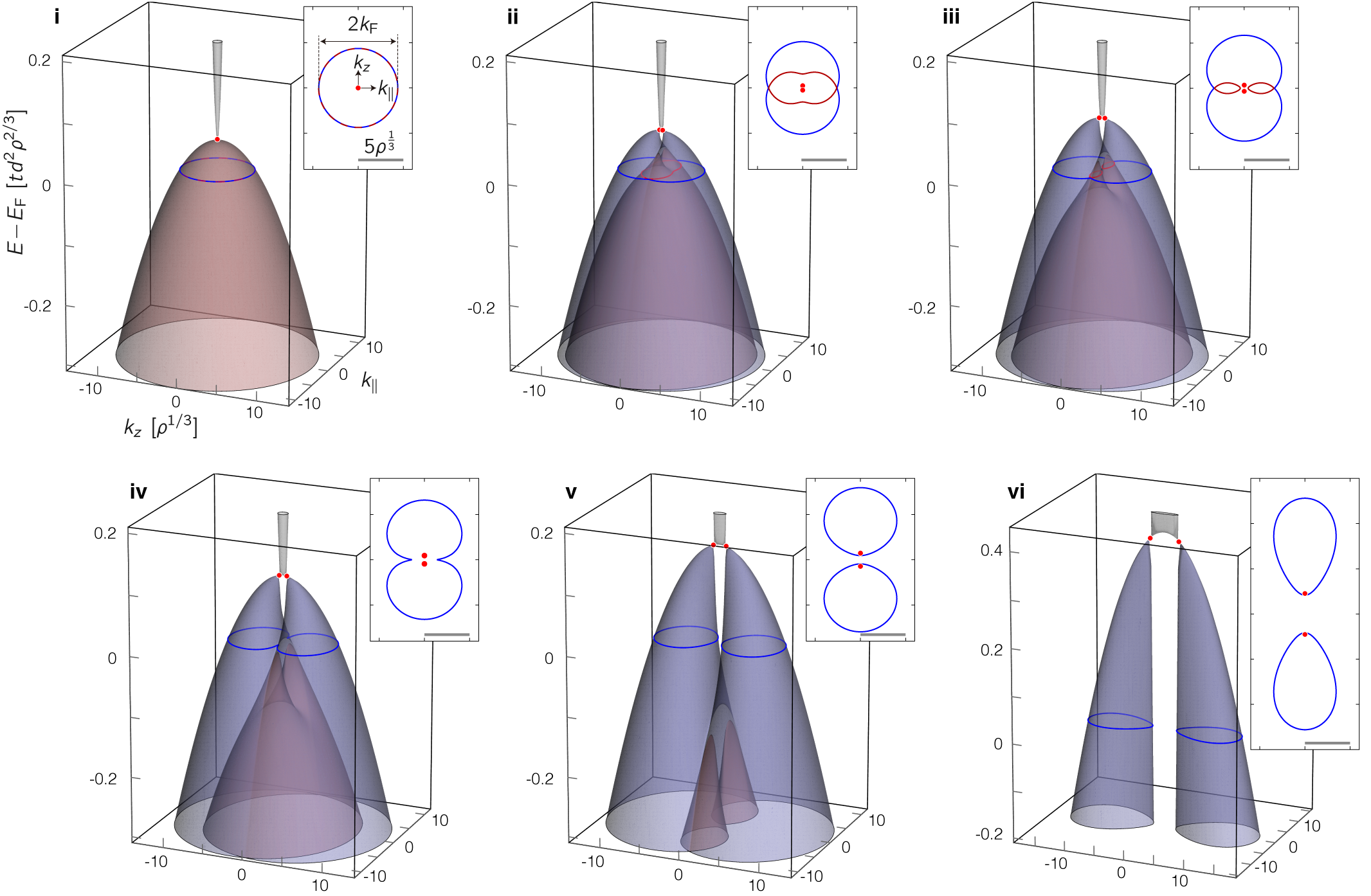}
	\caption{
	Evolution of the band structure with increasing magnetic field $B$ at fixed charge density $\rho$. Panels (i) to (vi) correspond to increasing values of $B$. The energy and wave vectors are shown in units independent of the magnetic field. Insets show cross sections of the Fermi surface in the $k_\parallel$-$k_z$ plane;   the full three-dimensional Fermi surface is obtained by rotating these profiles around the $k_z$-axis.  
    }\label{fig:fs}
\end{center}
\end{figure*}

In Fig.~\ref{fig:oscillation}(b), we present a Landau fan analysis for this system, where the index \( j \) of each oscillation peak is plotted as a function of the inverse magnetic field \( B^{-1} \).  
A notable feature, in contrast to conventional systems, is the deviation from perfect linearity.  
The nonlinearity is more clearly seen in Fig.~\ref{fig:lldisp}(c), which shows the spacing between neighboring peaks plotted against \( B^{-1} \). 
In the semiclassical framework, the spacing between peaks is related to the extremal area of the Fermi surface projected onto the plane perpendicular to the magnetic field, \( S^{\perp}_{\mathrm{F}} \), via the relation~\cite{ashcroftmermin}  
\begin{equation}
    B_{j+1}^{-1} - B_{j}^{-1} = \frac{2\pi e}{\hbar S_{\mathrm{F}}^{\perp}}.
\end{equation}  
In conventional systems, the Fermi surface remains unaffected by the magnetic field, resulting in  a constant peak spacing.

In contrast, the shape of the Fermi surface in the present system strongly depends on the magnetic field, resulting in unequally spaced peaks observed in Fig.~\ref{fig:lldisp}(c).  
The key factor underlying this behavior is the orbital Zeeman splitting.  
To illustrate this, we plot in Fig.~\ref{fig:fs} the band structure of $H_{\mathrm{eff+Z}}$ at different magnetic field strengths.  
The band structures are essentially identical to the universal dispersion in Fig.~\ref{fig:lldisp}(b), while the momentum and energy scales vary depending on the magnetic field.
The inset in each panel shows the Fermi surface on the $k_\parallel$--$k_z$ plane.  
The full 3D Fermi surface is obtained by rotating this 2D contour about the $k_z$-axis, and the maximum cross-sectional area $S_{\mathrm{F}}^{\perp}$ is extracted by projecting the 3D surface onto the $k_x$--$k_y$ plane.

As the magnetic field increases, the Zeeman splitting becomes more pronounced, and
the Fermi surface evolves from two degenerate spheres [Fig.~\ref{fig:fs}(i)] to split lobes along the $k_z$ direction [Fig.~\ref{fig:fs}(ii)]. 
With further increase in $B$, the Fermi surfaces become fully separated and elongated along the $k_z$ axis, leading to a reduced cross-sectional area in the $k_x$--$k_y$ plane [Fig.~\ref{fig:fs}(vi)]. 
This evolution directly influences the SdH peak spacing.  
In Fig.~\ref{fig:oscillation}(c), we
overlay the plot of $S_{\mathrm{F}}^{\perp}$ as a function of inverse magnetic field,
where we observe qualitative agreement with the spacing, evaluated from $(B_{n+1}^{-1} - B_n^{-1})$.
In addition to the overall nonlinear trends associated with Fermi surface evolution, 
two shoulder-like features appear at the vertical lines in Fig.~\ref{fig:oscillation}(c).
These correspond to the magnetic field $B$ where the Fermi energy intersects the minimum of the first Landau level [$\nu=+1$, yellow curve in Fig.~\ref{fig:lldisp}(a)] and the top of the second Landau levels ($\nu=-1$).
These results indicate that the characteristic Landau level structure of the singular flat bands can be identified from the non-uniform spacing of the SdH oscillations.

\section{Extention to High Angular Momentum flat bands }\label{sec:higherl}

The effective model in Eq.~\eqref{eq:heffl}, which describes the low-energy electronic states for the pyrochlore lattice, 
is based on orbital angular momentum $\ell=1$. 
This model can be generalized to higher angular momentum $\ell>1$ 
to explore broader class of flat band systems.
We define a generalized Hamiltonian of the form 
\begin{equation}\label{eq:hflat}
H_{\mathrm{eff}} = \frac{td^2}{2}\left(\ell^2\bm{k}^2-(\bm{k}\cdot\bm{L}^{(\ell)})^2\right).
\end{equation}
where $\bm{L}^{(\ell)}=(L_x^{(\ell)},L_y^{(\ell)},L_z^{(\ell)})$ are the set of angular momentum matrices for a given $\ell$, 
satisfying $(\bm{L}^{(\ell)})^2=\ell(\ell+1)$. The explicit matrix expression is provided in Appendix~\ref{sec:lmat}.
The eigenenergies are given by $E_{m}=(td^2/2)(\ell^2-m^2)k^2$, where $m=-\ell, -\ell+1, \cdots, \ell$ is the magnetic quantum number. 
Notably, the states with $m=\pm \ell$ become exactly flat with zero energy, $E_{\pm\ell}=0$, 
yielding a doubly degenerate flat band. 
The remaining $2\ell-1$ bands are dispersive and parabolic, touching the flat band from above.
The case of $\ell=1$ corresponds to the continuum model for the pyrochlore lattice introduced in Eq.~\eqref{eq:heffl}.
In addition, $\ell=3/2$ can be viewed as the Luttinger-Kohn model~\cite{luttinger1956}, $H_{\mathrm{LK}}= A\bm{k}^2 + B (\bm{k}\cdot\bm{L}^{(3/2)})^2$, with a special parameter choice $A=-(3/2)^2B$.


\begin{figure}
\begin{center}
\includegraphics[width=85mm]{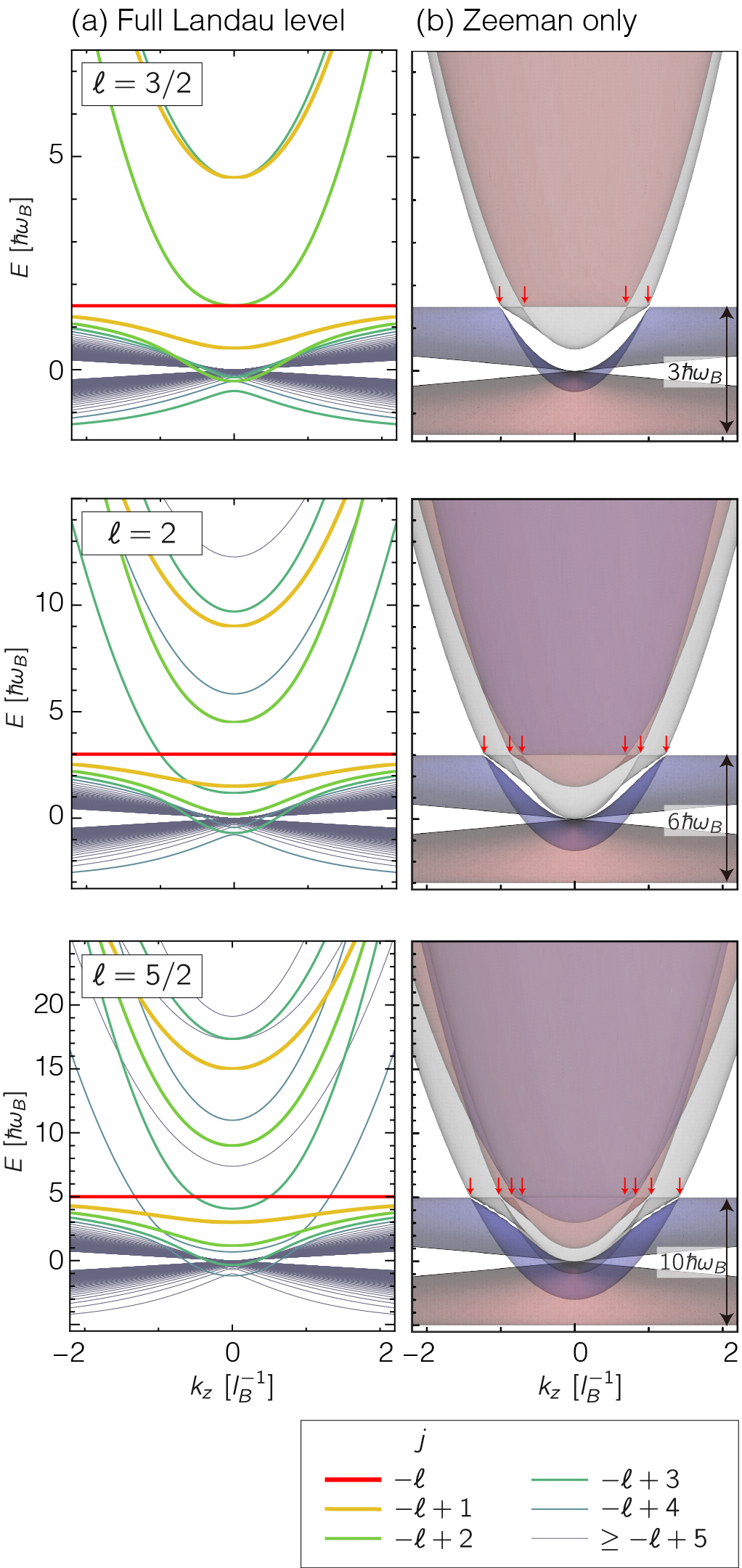}
	\caption{
	Magnetic field effect in higher angular momentum flat band systems. (a) Full Landau level spectrum. (b) Energy spectrum under the Zeeman term in continuum model. 
    Arrows indicate the position of the Weyl points. 
	}\label{fig:higherl}
\end{center}
\end{figure}

Within the same manner as Eq.~\eqref{eq:heffb}, the Landau quantization for general $\ell$ is formulated by replacing $\bm{k}$ with $\bm{\pi}/\hbar=-i\bm{\nabla}+(e/\hbar)\bm{A}$. 
The resulting Hamiltonian in the presence of a magnetic field is given by
\begin{align}\label{eq:heffbl}
	H_{\mathrm{B}}(k_z) = \frac{\hbar\omega_B}{2} \Bigg[\Big(\ell(\ell-1)+(L_z^{(\ell)})^2\Big)\left(a^\dag a + \frac{1}{2}\right) 
    \nn\\+ \Big(\ell^2-(L_z^{(\ell)})^2\Big) (k_z l_B)^2 \nn\\
		-   \frac{L_-^{(\ell)}}{\sqrt{2}}(2L_z^{(\ell)}-1)   k_z l_B a^\dag-   \frac{L_+^{(\ell)}}{\sqrt{2}}(2L_z^{(\ell)}+1) k_z l_B a\nn\\
		- \frac{(L_-^{(\ell)})^2}{2}a^{\dag2} -  \frac{(L_+^{(\ell)})^2}{2}a^2 \Bigg]
\end{align}
We adopt the basis given by Eq.~\eqref{eq:basis}, generalized to a magnetic quantum number $m=-\ell,\cdots,\ell$.
The total angular momentum
$j=n+m=-\ell, -\ell+1, \cdots$, is defined as the eigenvalue of the operator $J_z=a^\dag a + L^{(\ell)}_z$, and remains conserved due to the commutation relations. 
The Hamiltonian for each sector labeled by $j\ge\ell$ takes the form
\begin{equation}\label{eq:hllhigher}
H_{B}(j,k_z) = \tilde{H}_B(j,k_z)+H_{\mathrm{Z}},
\end{equation}
which is a matrix of dimension $2\ell+1$. 
Here $\tilde{H}_B(j,k_z)$ is structured as 
\begin{equation}\label{eq:btill}
\tilde{H}_{B}=\frac{\hbar\omega_B}{2}\left(\begin{array}{cccccccccccc}
    D_\ell & U_{\ell-1} & V_{\ell-2} & 0 & \cdots & 0 \\
    U_{\ell-1} & D_{\ell-1} & U_{\ell-2} & V_{\ell-3} & \ddots & \vdots\\
    V_{\ell-2} & U_{\ell-2} & D_{\ell-2} & U_{\ell-3}  & \ddots & 0 \\ 
    0& V_{\ell-3} & U_{\ell-3} & D_{\ell-3}  &\ddots & V_{-\ell}\\
    \vdots&\ddots&\ddots&\ddots& \ddots & U_{-\ell}\\
    0&\cdots&0&V_{-\ell}&U_{-\ell}& D_{-\ell}
\end{array}\right),
\end{equation}
with matrix elements explicitly given by
\begin{align}
D_{m}=&\Big(\ell(\ell-1)+m^2\Big)(j-m+1)\nn\\
&-\frac{1}{2}(\ell-m)(\ell-m-1) +(\ell^2-m^2)k_z^2 l_B^2 \\
U_{m}=& -\frac{(2m+1)k_z l_B}{\sqrt{2}}\sqrt{(j\!-\!m)(\ell\!-\!m)(\ell\!+\!m\!+\!1)} \\
V_{m}=& \frac{1}{2}\sqrt{(j\!-\!m)(\ell\!-\!m)(\ell\!+\!m\!+\!2)} \nn\\
&\times \sqrt{(j\!-\!m\!-\!1)(\ell\!-\!m\!-\!1)(\ell\!+\!m\!+\!1)}.
\end{align}
The Zeeman term 
\begin{equation}\label{eq:hzl}
H_{\mathrm{Z}} = \frac{\hbar\omega_B}{2}\left(\ell-\frac{1}{2}\right)L_{z}^{(\ell)},
\end{equation}
with $L_z=\mathrm{diag}(\ell,\ell-1,\cdots,-\ell)$ is naturally incorporated and leads to a splitting of levels according to the angular momentum projection $m
$. 
For sectors with $-\ell\le j<\ell$, 
the projected Hamiltonian is obtained as the lower-right $(\ell+j+1)\times(\ell+j+1)$ submatrix of 
$H_{B}(j,k_z)$, following the same construction as described in Sec.~\ref{sec:ll}.

Figure~\ref{fig:higherl}(a) shows the energy spectrum for the Landau level Hamiltonian $H_{B}$ for $\ell=3/2, 2$ and 5/2.
The Landau levels originating from the flat band are spread over the range $-\Delta_Z < E < \Delta_Z$, where
\begin{equation}
\Delta_{\mathrm{Z}} = \frac{\hbar \omega_B}{2} \ell \left( \ell - \frac{1}{2} \right),
\end{equation}
indicating that the spread becomes wider with increasing $\ell$.
The $j=-\ell$ level, shown as  a red line, remains perfectly flat as a function of $k_z$,
and also isolated from the main cluster the Landau levels.
Notably, for arbitrary $\ell$, the Landau level spreading in the flat band is completely suppressed in the reduced Hamiltonian $\tilde{H}_{B} = H_{B} - H_{\mathrm{Z}}$, as already observed in the case of $\ell = 1$ [Fig.~\ref{fig:llflat}(b)].
This can be shown by noting that the reduced Hamiltonian $\tilde{H}_{B}$ [Eq.~\eqref{eq:btill}] has rank $2\ell - 1$, which is two less than its full dimension $2\ell + 1$, implying the existence of two zero modes. We thus conclude that, regardless of the angular momentum $\ell$, the Landau level spreading originates solely from the Zeeman term $H_{\mathrm{Z}}$.

\begin{figure}[b]
\begin{center}
	\includegraphics[width=85mm]{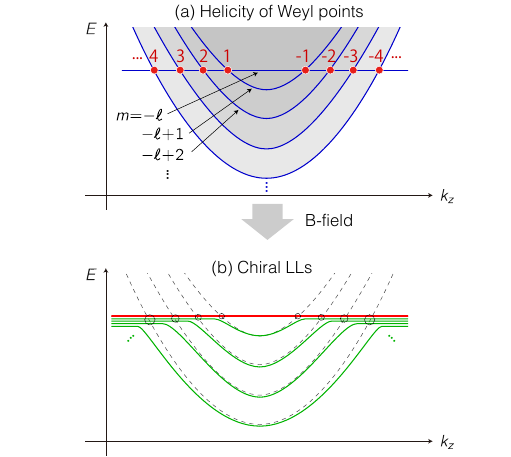}
	\caption{
        (a) Schematic diagram illustrating a series of Weyl points and their helicities in a flat band system with higher angular momentum.
        (b) Corresponding chiral Landau levels.
        Here $\chi$ branches penetrate the Weyl point with helicity $\chi$ shown in panel (a).}
        \label{fig:schematic_highell}
\end{center}
\end{figure}

The Landau level spectrum reflects the band structure of the effective model with the Zeeman term, $H_{\mathrm{eff}+Z} = H_{\mathrm{eff}} + H_{\mathrm{Z}}$, which is shown in Fig.~\ref{fig:higherl}(b). 
Importantly, there are $2\ell-1$ Weyl points on the $k_z$ axis, which correspond to the band inversion of the states with different angular momentum $m$, as schematically shown in Fig.~\ref{fig:schematic_highell}(a). 
As $k_z$ increases, the $m=-\ell$ band (the flat band) crosses the $m=-\ell+1$, $-\ell+2$, $\cdots$, $\ell-1$ bands. 
\blue{Near each crossing point, the low energy physics is described by a two-band effective model involving the $m=-\ell$ and $m'$ states. The corresponding Weyl Hamiltonian takes the form 
\begin{equation}
    H_\mathrm{W}= \Delta_Z + 
    \left(\begin{array}{cc}
    0 & A k_+^{\ell+m'} \\
    A k_-^{\ell+m'} & v_{m'}(k_z-k_\mathrm{W})
    \end{array}\right)
\end{equation}
where $k_{+} = k_x + ik_y$, $A$ is a real parameter and $v_{m'}$ is the group velocity of of the $m'$ state along the $z$ direction at the Weyl point at $\bm{k}=k_{\mathrm{W}} \hat{\bm{z}}$.
In this system, velocity satisfies $v_{m'} k_{\mathrm{W}}>0$.
Consequently, each band inversion between $m=-\ell$ and $m'$ yields a Weyl point with helicity $\chi=\pm(m-m')$ depending on whether $k_{\mathrm{W}}\gtrless0$. 
The system therefore hosts Weyl points with $\chi=\pm 1$, $\pm 2$, $\pm 3$, $\cdots$ $\pm(2\ell-1)$, where each helicity corresponds to the Chern number defined on a sphere enclosing the respective Weyl point.}

Under the magnetic field, each Weyl points generates chiral Landau levels, 
where $\chi$ levels penetrate the Weyl point of helicity $\chi$, as schematically shown in Fig.~\ref{fig:schematic_highell}(b).
Here, we are left with a single Landau level that penetrates all the Weyl points, corresponding to the perfectly flat Landau level with $j = -\ell$ shown in Fig.~\ref{fig:higherl}(a).
We note that the Landau level branching structures as shown in Fig.~\ref{fig:schematic_highell}(b) becomes more evident if we artificially enhance the Zeeman splitting $H_{\mathrm{Z}}$ compared to the Landau level spacing, as shown in Appendix~\ref{sec:etahz}.

\section{Summary}
We have developed a theoretical framework for three-dimensional singular flat band systems, 
with a particular focus on the pyrochlore lattice,
that hosts a band touching between a doubly degenerate flat band and a dispersive band. 
By deriving a three-orbital effective continuum model, 
we revealed a point-like topological singularity at the band-touching point. 
This singularity governs the quantum geometry of the flat band manifold and 
plays a central role in its magnetic response.

We identified a characteristic $k_z$-dependent Landau level structure under a magnetic field, which reflects a continuous evolution of the quantum geometry from a trivial regime at large $|k_z|$ to a singular structure near $k_z=0$.
Such behavior is a distinctive feature of three-dimensional singular flat band systems.
%
%
We further demonstrate that the entire Landau level structure can be well described by
an effective band model including the orbital Zeeman effect. 
This approach uncovers a Weyl-semimetal-like response emerging form the flat bands with the Zeeman effect. It also provides a framework for analyzing quantum geometric effects in degenerate flat bands, by considering the quantum geometric tensor in each of the Zeeman-split bands.
Specifically, we demonstrated that the Landau level spreading  in the large $|k_z|$ regime 
is proportional to the quantum geometric tensor in the split band.

In addition, we investigated the effect of a weak dispersion induced by longer-range hopping.
The interplay between this dispersion and orbital magnetism leads to unconventional Shubnikov–de Haas oscillations with a slowly drifting period, 
originating from the field-induced evolution of the Fermi surface 
driven by Zeeman effect. Finally, we extended our continuum framework to arbitrary orbital angular momentum $\ell \ge 1$, confirming the robustness of these phenomena across a broad class of three-dimensional flat band systems.

\begin{acknowledgements}
This work was supported by JSPS KAKENHI Grants No. JP20K14415, No. JP20H01840, No. JP20H00127, No. JP21H05236, No. JP21H05232, JP24K06921, JP25K00938, and by JST CREST Grant No. JPMJCR20T3, Japan.
\end{acknowledgements}

\appendix
\section{Asymmetric forms of Landau levels in pyrochlore model}
\subsection{$k_z\rightarrow \infty$ regime}\label{sec:drv_asym}
Here we provide a perturbative analysis of the Landau level spectrum leading to Eq.~\eqref{eq:elargek}, 
by taking advantage of the fact that the eigenvalue problem 
\begin{equation}\label{eq:ephbtil}
\tilde{H}_{B}(j,k_z)\vec{\Phi}_{\nu}=\tilde{E}_{\nu}(j,k_z) \vec{\Phi}_{\nu}
\end{equation}
for the Zeeman-term-subtracted Hamiltonian $\tilde{H}_{B}(j,k_z)=H_{\mathrm{B}}(j,k_z)-H_{\mathrm{Z}}$, introduced in Eq.~\eqref{eq:hj},  can be solved analytically. 
The corresponding eigenvalues and eigenvectors of Eq.~\eqref{eq:ephbtil} are given by
\begin{equation}
	\tilde{E}_{\pm 1}(j,k_z)  = 0,\quad \tilde{E}_{0}(j,k_z) = \hbar\omega_B \lambda,
\end{equation}
with 
\begin{equation}
	\lambda = \lambda(j,k_z) = j+\frac{1}{2}+\frac{k_z^2l_B^2}{2}
\end{equation}
and
\begin{gather}
	\vec{\Phi}_{-1} = \frac{1}{\sqrt{2j+1}}
	\left(\begin{array}{c}
	\sqrt{j+1} \\
	0 \\
	\sqrt{j} 
	\end{array}\right)\nn\\
	\vec{\Phi}_{+1} = \frac{1}{\sqrt{2\lambda_j(2j+1)}}
	\left(\begin{array}{c}
	\sqrt{j}k_zl_B \\
	2j+1\\
	-\sqrt{j+1}k_zl_B 
	\end{array}\right)\nn\\
	\vec{\Phi}_{0} = \frac{1}{\sqrt{2\lambda_j}}
	\left(\begin{array}{c}
	-\sqrt{j} \\
	k_z l_B\\
	\sqrt{j+1} 
	\end{array}\right).
\end{gather}
Note that $\tilde{E}_{\pm 1}(j,k_z)$ are independent of both  $k_z$ and the total angular momentum $j$, and thus represent flat bands.  
In contrast, $\tilde{E}_{0}(j,k_z)$ corresponds to the conventional Landau levels  
for a three dimensional free electron.

We now consider the perturbative analysis 
by treating $\tilde{H}_B(j,k_z)$ as the unperturbed Hamiltonian and 
$H_{\mathrm{Z}}$ as a perturbation, 
assuming the regime $\lambda_j\gg1$.
In the basis of eigenstates $\vec{\Phi}_{\nu}$ of $\tilde{H}_B$, the total Hamiltonian is written as
\begin{equation}
	H_{B}(j,k_z) = 
	\left(\begin{array}{cc}
	F & U^\dag \\
	U & D
	\end{array} \right)
\end{equation}
where $[F]_{s,s'}=\vec{\Phi}_{s}^\dag H_{\mathrm{Z}}\vec{\Phi}_{s'}$ with $s,s'= \pm 1$ 
and $D=\hbar\omega_B \lambda + \vec{\Phi}_{0}^\dag H_{\mathrm{Z}}\vec{\Phi}_{0}$ 
represent the blocks of flat and dispersive bands, 
and $U=(\vec{\Phi}_{0}^\dag H_{\mathrm{Z}} \vec{\Phi}_{-1},\vec{\Phi}_{0}^\dag H_{\mathrm{Z}} \vec{\Phi}_{+1})$ 
is coupling between two sectors.
For $\lambda\gg1$, the two blocks are energetically separated, 
and the Hamiltonian in the $s = \pm 1$ sector describes the low-energy physics.
Applying standard second-order perturbation theory, 
the projected Hamiltonian in this sector is given by
\begin{equation}
	H_{\mathrm{proj}} = F - U^\dag D^{-1} U.
\end{equation}
Solving this $2\times 2$ Hamiltonian, we obtain the energy spectrum 
\begin{align}\label{eq:analytic}
	&E_{\pm1}(j,k_z)\sim \frac{\hbar \omega_B}{16(2j+1)} \Bigg(-2+\frac{k_z^2 l_B^2}{\lambda} + \frac{j(j+1)}{\lambda^2} \nn\\
	&\pm \sqrt{\left(2+\frac{k_z^2 l_B^2}{\lambda} - \frac{j(j+1)}{\lambda^2}\right)^2 + \frac{32 k^2_z l_B^2 j(j+1)}{\lambda}}\Bigg). 
\end{align}
Here, we use the estimations $k_z\l_B \lesssim \sqrt{\lambda}$ and $j \lesssim \lambda$, retaining terms up to $\lambda^{-1}$ while
neglecting higher-order correction of $\mathcal{O}(\lambda^{-2})$ .
The expression Eq.~\eqref{eq:analytic} captures the asymptotic behavior in Fig.~\ref{fig:llflat}. 
In the more restrictive limit $j\ll k_z l_B$, the expression reduces to Eq.~\eqref{eq:elargek}.%

\subsection{$k_z\sim 0$ regime}
For $k_z=0$, the eigenenergies for $j\ge 1$ are given by
\begin{equation}
	E_{0}^{(j)}(0) = 0,\quad E_{\pm 1}^{(j)}(0) = \frac{\hbar\omega_B}{2} \left(j+\frac{1}{2}\pm \lambda \right)
\end{equation}
with
\begin{equation}
	\lambda(j) = \sqrt{j(j+1)+1}
\end{equation}
The corresponding wave functions are 
\begin{gather}
	\vec{\Psi}_{\pm 1,0}^{(j)} = 
	\frac{1}{\sqrt{2\lambda(\lambda\pm 1)}}
	\left(\begin{array}{c}
		\sqrt{j(j+1)}\\
		0\\
		-1\mp \lambda
	\end{array}
	\right), \ 
	\vec{\Psi}_{0,0}^{(j)} = 
	\left(\begin{array}{c}
		0\\
		1\\
		0
	\end{array}
	\right).
\end{gather}
The eigenenergy at finite momentum $k_z\sim0$ is approximated 
using standard perturbation theory as, 
\begin{align}
	E_\nu^{(j)}(k_z) \simeq \langle{\Psi}_{\nu,0}^{(j)}| H_B^{(j)}|\Psi_{\nu,0}^{(j)}\rangle 
	-\sum_{\nu'\neq \nu} \frac{|\langle\Psi_{\nu',0}^{(j)} |H_B^{(j)}|\Psi_{\nu,0}^{(j)}\rangle|^2}{E_{\nu'}(0)-E_{\nu}(0)}
\end{align}
To the second order of the momentum $k_z$, we find
\begin{gather}
	E_{\pm 1}^{(j)}(k_z) \simeq E_{\pm 1}^{(j)}(0) - \frac{\hbar{\omega_B}(1+j)(j+1\pm\lambda)^2}{4\lambda(\lambda\pm 1)(j+\frac{1}{2}\pm\lambda)}k_z^2 l_B^2\\
	E_{0}^{(j)}(k_z) \simeq  \frac{\hbar{\omega_B}}{6}k_z^2 l_B^2.
\end{gather}
These results well reproduce the numerically obtained spectrum near $k_z\sim0$ (Fig.~\ref{fig:llflat}).

\section{Zeeman splitting in Kagome Lattice model}
Since the effective continuum Hamiltonian in Eq.~\eqref{eq:heff} naturally extends the Kagome lattice model to three-dimensions, 
the explanation for the Landau level spreading in terms of the orbital Zeeman splitting can be directly applied to the Kagome lattice as well. 

The effective continuum Hamiltonian for the Kagome lattice is given by~\cite{rhim2020}
\begin{equation}\label{eq:hkag}
    H_{\mathrm{Kag}}(\bm{k})=\frac{td^2}{2}
    \left(\begin{array}{cc}
        k_x^2 & k_x k_y \\
        k_x k_y & k_y^2
    \end{array}\right)
\end{equation}
Using the rotational symmetry about $z$-axis, $R_z(\theta)H_{\mathrm{Kag}}(\bm{k})R_z^{-1}(\theta)=H_{\mathrm{Kag}}(R_z(\theta)\bm{k})$, 
we define the orbital angular momentum operator
\begin{equation}\label{eq:lzkag}
L_{z} = 
\left(\begin{array}{cc}
0 & -i \\
i & 0 
\end{array}\right)
\end{equation}
within the basis of Eq.~\eqref{eq:hkag}. 
By transforming the basis to that which diagonalize $L_z$ in Eq.~\eqref{eq:lzkag}, i.e., 
\begin{equation}
L_z \rightarrow L'_z = \left(\begin{array}{cc}
1 & 0 \\
0 & -1 
\end{array}\right)
\end{equation}
and applying the ladder operator substitution, 
the Landau level Hamiltonian is obtained as 
\begin{equation}
    H_{\mathrm{B}}^{(\mathrm{Kag})}=\frac{\hbar\omega_B}{2}
    \left(
    \begin{array}{cc}
    a^\dag a + \frac{1}{2} & \frac{1}{2}a^2 \\[5pt]
    \frac{1}{2}a^{\dag 2} & a^\dag a + \frac{1}{2}
    \end{array}\right).
\end{equation}
This Hamiltonian commute with 
\begin{equation}
    J^{(\mathrm{Kag})}_z=a^\dag a + L'_z, 
\end{equation}
so the eigenvalue $j>-1$ of $J^{(\mathrm{Kag})}_z$ is good quantum number. 
The projected Hamiltonian in $j$-sector for $j\ge0$ is given by
\begin{equation}\label{eq:hllkag}
    H_{B}^{(\mathrm{Kag},j)} = \frac{\hbar\omega_B}{2}
    \left(\begin{array}{cc}
    j & -\sqrt{j(j+1)} \\
    -\sqrt{j(j+1)} & j + 1
    \end{array}\right) - \frac{\hbar\omega_B}{4}L'_z.
\end{equation}
For $j=-1$ sector, the Hamiltonian is represented by the lower-right element of $H_{LL}^{(\mathrm{Kag},-1)}$, yielding a Landau level at $E=\hbar\omega_B/4$.
The first term in Eq.~\eqref{eq:hllkag} clearly gives eigenvalues
\begin{equation}
\epsilon = \hbar\omega\left(j+\frac{1}{2}\right) \hbox{ and } 0
\end{equation}
Corresponding to the Landau levels of the trivial parabolic band 
and the flat band.
However, due to the second term in \eqref{eq:hllkag}, 
which accounts for the orbital Zeeman splitting, 
the resulting Landau level takes the form
\begin{equation}
    E_{\pm}=\frac{\hbar\omega_B}{2}\left(j+\frac{1}{2}\pm\sqrt{j(j+1)+1}\right)
\end{equation}
with having eigenvalue $E_-<0$, where no continuum states exists. 
This spreading, which is related to the quantum geometric effect~\cite{rhim2020}, can also be understood in terms of the orbital Zeeman effects.

\section{Formulating the density of states under the magnetic field}\label{sec:dosderive}
To analyze the SdH oscillations in Sec.~\ref{sec:sdh}, we formulate the charge density and the density of states
within the present Landau-level Hamiltonian, 
incorporating the appropriate cutoffs and normalization. 
Here we assume the pyrochlore lattice model, with the Brillouin zone boundary at $k_z=\pi/2d$. 
When the Landau levels below a fixed energy $E$ are occupied, the charge density is given by
\begin{align}\label{eq:rho}
    \rho(E,B) =\!\!\!\! \sum_{\substack{\nu,j\\n<n_c}} \!\!\int_{-\frac{\pi}{2d}}^{\frac{\pi}{2d}}\frac{dk_z}{{2\pi \Lambda^2}} g_B f(E_{\nu}(j,k_z)-E)-1/d^3 
\end{align}
where $\Lambda$ represents the system size and 
the Landau level degeneracy is given by $g_{B}=\Lambda^2/l_{B}^2$. 
The occupation of the states is determined by the Fermi distribution function $f(E)=(1+e^{\beta E})^{-1}$, which is centered at the Fermi energy. 
We subtract $1/d^3$ in Eq.~\eqref{eq:rho} 
to define the zero point of the charge density, $\rho=0$ to correspond to a fully filled flat band, in which two electrons occupy each unit cell of volume $2d^3$ (see Fig.~\ref{fig:pyrochlore_lattice}(a)). 
To ensure that the total state density, accounting for three orbitals per unit cell, satisfies the condition $\lim_{E\rightarrow \infty}\rho(E,B)=1/(2d^3)$, 
we introduce a cutoff for the Landau level index 
$n=j-m$ as $n_c = l_B^2/d^2$. 
Using the charge density from Eq.~\eqref{eq:rho}, 
the density of states at the Fermi energy is given by 
\begin{equation}
    D_{\mathrm{F}}(\rho,B) = \frac{\partial \rho}{\partial E}\Big|_{E=E_{\mathrm{F}}(\rho,B)}
\end{equation}

We now introduce the formulation in magnetic units.
The energy, momentum, and Fermi distribution are normalized as 
$\tilde{E}=E/\hbar\omega_B$, 
$\tilde{k} = k_z l_B$, and 
$\tilde{f}(\tilde{E})=(1+e^{\tilde{\beta}\tilde{E}})^{-1}$ 
where the normalized inverse temperature is given by 
$\tilde{\beta}=\beta\hbar\omega_B$.
Within this parametrization, 
the charge density, normalized as 
$\tilde{\rho}=\rho l_B^3$, is given by 
\begin{equation}\label{eq:p}
\tilde{\rho}(\tilde{E},\gamma) = \frac{1}{2\pi}\sum_{\substack{\nu,j\\n<\gamma^{-2}}} \int_{-\frac{\pi}{2\gamma}}^{\frac{\pi} {2 \gamma}}d\tilde{k}\tilde{f}(\tilde{E}_{\nu}^{(j)}-\tilde{E})-\gamma^{-3}.
\end{equation}
with $\gamma(B) = d/l_B$. 
Using this normalized charge density, the density of states is expressed as 
\begin{equation}\label{eq:dapp}
    D_{\mathrm{F}}(\rho,B) = \frac{\gamma}{td^3}\frac{\partial \tilde{\rho}}{\partial \tilde{E}}\bigg|_{\tilde{E}=\tilde{E}_{\mathrm{F}}(\tilde{\rho},\gamma)}
\end{equation}

\begin{figure}[t]
\begin{center}
	\includegraphics[width=85mm]{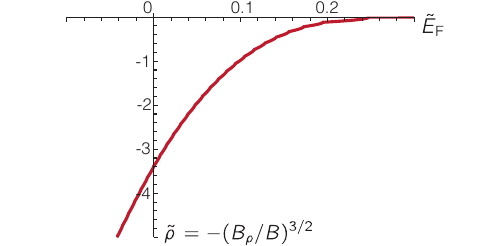}
	\caption{
	Numerically obtained normalized charge density $\tilde{\rho}-\tilde{\rho}_0=\rho l_{\mathrm{B}}^3-\gamma^{-3}=(B/B_{\rho})^{3/2}$ as a function of normalized Fermi energy $\tilde{E}_{F}=E_{\mathrm{F}}/\hbar\omega_B$ and }
	\label{fig:efp}
\end{center}
\end{figure}

Specifically, we consider the regime where the Fermi energy lies slightly below the top of the bent flat band and $t'>0$, corresponding to $E_F\geq E_{+1}(-1,\pi/2d)$.
In this regime, we find the ordinary scaling behavior of the density of states $D_{\mathrm{F}}$ with respect to $\rho$ and $B$.
The key factor is that the normalized charge density $\tilde{\rho}$ given in Eq.~\eqref{eq:p} is independent of $\gamma$ and is determined solely by $\tilde{E}_{\mathrm{F}}$. 
Conversely, $\tilde{E}_{\mathrm{F}}$ is uniquely determined by $\tilde{\rho}$.
The typical functional dependence between $\tilde{\rho}$ and $\tilde{E}_F$ is given in Fig.~\ref{fig:efp}.
When we introduce a characteristic scale for the magnetic field, given by
\begin{equation}
    B_{\rho} = \frac{\hbar}{e}|\rho|^{2/3},
\end{equation}
normalized Fermi energy in Eq.~\eqref{eq:dapp} is given by $\tilde{E}_{\mathrm{F}}=\tilde{E}_{\mathrm{F}}(B/B_\rho)$
Therefore, the normalized density of states is expressed as
\begin{equation}\label{eq:dtil}
    \tilde{D}_{\mathrm{F}}(B/B_{\rho}) = \frac{td^2D_F}{|\rho|^{1/3}}
    =\sqrt{\frac{B}{B_{\rho}}}\frac{\partial \tilde{\rho}}{\partial \tilde{E}}\bigg|_{\tilde{E}=\tilde{E}_{\mathrm{F}}(B/B_\rho)} 
\end{equation}

\section{Divergent magnetic response in the flat-band limit}
It has recently been theroetically proposed that 
the two dimensional singular flat bands exhibit 
the logalithmically divergent magnetic susceptivility, 
which is associated with the spread of their Landau levels.
Here, a natural question arises: 
what about the magnetic suceptivility of the three dimensional singular flat bands? 
This is the topic we discuss in this section.

\begin{figure}[t]
\begin{center}
	\includegraphics[width=85mm]{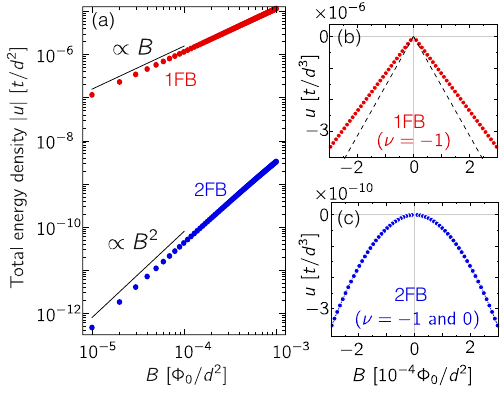}
	\caption{
	Numerically obatained field dependence of the total energy density $u$ (Eq.~\eqref{eq:dimlessu}), 
	blue and red dots in log-log plot (a) are the result
	for different fillings: $\nu=1$ where the half filling of 
	twofold degenerate flat band, 
	$\nu=2$ where the two flat bands are fully filled.
	(b) and (c) is the linear scale plot for the two cases. 
	}\label{fig:utot}
\end{center}
\end{figure}

In the same way as the charge density, 
the normalized total energy density $\tilde{u}=u/(t/d^3)$ is derived as 
\begin{align}\label{eq:dimlessu}
	  \tilde{u} = \frac{\gamma^5}{2\pi}  \sum_{\substack{\nu,j\\n<1/\gamma^{2}}} \int_{-\frac{\pi}{2\gamma}}^{\frac{\pi}{2\gamma}} d\tilde{k} \tilde{E}_{\nu}^{(j)}(\tilde{k}) \tilde{f}(\tilde{E}_{\nu}^{(j)}(\tilde{k})-\tilde{E}_F)
\end{align}
We consider the case with perfect flat band, as shown in Fig.~\ref{fig:llflat} and numerically evaluate the low-field total energy $u$ in Eq.~\eqref{eq:dimlessu} as a function of the field $B$, 
as presented in Fig.~\ref{fig:utot}. 
When the twofold degenerate flat bands are half filled, 
that is, for the charge density $\rho=1/d^{-3}$, where $\nu=-1$ is filled and $\nu=0$ and $\nu=1$ are unoccupied (see Fig.~\ref{fig:llflat}(a)). 
The total energy $u$ increases linearly with the magnetic field $B$. 
At $B=0$ since the slope of total energy $\tilde{u}$ undergoes discontinuous change, 
the suceptibility exhibits a delta-function-like behavior,
\begin{equation}
	\chi=-\frac{\partial^2 u}{\partial B^2} \sim \delta(B),
\end{equation}
indicating the the divergent magnetic response.
This behavior can be understood as follows. 
As shown in the Landau level in Fig~\ref{fig:llflat}(a), 
and also analytically derived in Eq.~\eqref{eq:elargek}, 
for $k_z l_B \gg j$, the spectrum splits into $E=\pm \hbar\omega_B/4$ 
due to the sublattice orbital Zeeman effect.
The region $k_z l_B \gg j$ can be regarded as an isolated two-level system. 
If we assume $E_{\nu}^{(j)}(k_z)=\pm\hbar\omega_B/4$ as the simplest approximation 
and the Fermi energy is located at $E=0$,
Eq.~\eqref{eq:dimlessu} gives
\begin{equation}\label{eq:analyticu}
	\tilde{u} = \gamma_B^2/2^{6} \propto  B
\end{equation}
which qualitatively reproduces the linear behavior in Fig.~\ref{fig:utot} (b).
The solution Eq.~\eqref{eq:analyticu} is plotted as the dashed line in Fig.~\ref{fig:utot} (b) 
which shows the order of the slope of $\tilde{u}$.
However, analysis quantitatively overestimate the slope of $u-B$ curve.
This is because around $j\sim n_{c}$ 
in the integration interval $|k_zl_B| \lesssim \sqrt{n_c}$ 
the region where asymptotic behavior in Eq.~\eqref{eq:elargek} 
is not relevant becmes non-negligible.
Nontheless, the linear dependence of $\tilde{u}$ remains robust,
as confirmed by the full numerical calculation.

Let us consider the case with two flat bands are occupied in the zero-field system. 
Under the applied magnetic field, Landau levels of $\nu=-1$ and $\nu=0$ in Fig.~\ref{fig:llflat}(a)
are both occupied and Fermi energy is slightly above $E=\hbar\omega_B/4$.
Energy increase in $\nu=0$ and the energy decrease in $\nu=-1$ cancels each other.
However, as shown in the numerical calculation in Fig.~\ref{fig:utot}(c) 
a finite negative energy remains, exhibiting paramagnetism. 
This behavior is purely a consequence of the Landau level structure around $k_zl_B\lesssim 1$ in Fig.~\ref{fig:llflat}.

\section{Matrix Representation of Angular momentum} \label{sec:lmat}
The angular momentum with arbitrary angular momentum $\ell$ is described as a Hermitian operator $\bm{L}=(L_x,L_y,L_z)$ that satsifies the commutation relation
\begin{align}\label{eq:comm}
	[L_i,L_j]=i \epsilon_{ijk}L_k.
\end{align}
where $\epsilon_{ijk}$ is the Levi-Civita symbol. 
The eigenstates of $\bm{L}^2$ and $L_z$ are denoted by $|\ell,m\rangle$ and satsfy 
\begin{align}
	\bm{L}^2| \ell, m \rangle &= \ell(\ell+1)| \ell, m\rangle, \\ 
	L_z| \ell, m \rangle & = m | \ell, m\rangle.
\end{align}
with $\ell$ being a non-negative integer or half-integer and $m=\ell, \ell-1,\cdots,-\ell$.
Here, $\ell$ lavels the total angular momentum, and $m$ gives its projection along $z$-axis. 

The operators $L_x$ and $L_y$ do not commute with $L_z$, but their matrix elements can be evaluated using the laddder operators 
\begin{equation}
    L_{\pm}=L_x\pm i L_y
\end{equation}
which raise or lower the $m$ quantum number by one unit. 
These operators act on the basis states as 
\begin{equation}
    L_{\pm}|\ell,m\rangle = \sqrt{(\ell\mp m)(\ell\pm m+1)}|\ell,m\pm 1\rangle
\end{equation}
From this, the matrix representations of the angular momentum operators $L_x$, $L_y$ and $L_z$ can be derived explicitly: 
\begin{align}
	\langle \ell, m | L_x | \ell, m' \rangle &=
		\frac{1}{2} \sqrt{(\ell\pm m)(\ell\mp m+1)} \delta_{m,m'\pm 1} \\
	\langle \ell, m | L_y | \ell, m' \rangle &= 
		\mp \frac{i}{2} \sqrt{(\ell\pm m)(\ell\mp m+1)} \delta_{m,m'\pm 1} \\
	\langle \ell, m | L_z | \ell, m' \rangle &= 
		 m \delta_{m,m'}
\end{align}
respectively.
\begin{figure}[t]
\begin{center}
	\includegraphics[width=85mm]{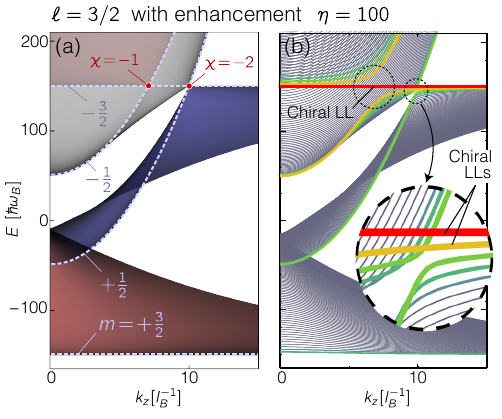}
	\caption{
        Energy spectrum for the continuum Hamiltonian (a) and Landau-level Hamiltonian (b) with a Zeeman field artificially enhanced by a factor of $\eta$, for $\ell=3/2$. Panel (a) corresponds to a scaled version of Fig.~\ref{fig:higherl}(b), featuring Weyl points with helicities $\chi=1$ and $2$. 
        The dashed curves in (a) show the dispersion along high-symmetry line $k_x=k_y=0$. In panel (b), the Landau level structure characteristic to the Weyl points is indicated by dahsed circle.
	}\label{fig:enhanced32}
\end{center}
\end{figure}

\section{High-angular momentum model with enhanced Zeeman term}\label{sec:etahz}
To clarify the correspondence between picture of the chiral Landau levels shown in Fig.~\ref{fig:schematic_highell} 
and real Landau level Fig.~\ref{fig:higherl}(a), 
we consider a modified effective Hamiltonian in which the Zeeman term is artificaially enhanced. 
This model is defined by Eq.~\eqref{eq:effetaz} and \eqref{eq:beta} along with Eq.~\eqref{eq:hflat}, \eqref{eq:hllhigher},  and 
\eqref{eq:btill}. 
While the enhanced Zeeman term increases the energy splitting, 
it does not change the topological properties of the system. 
Instead, it reduces the Landau level spacing relative to the separation of the Weyl points. 
This allows for a clearler comparison between discrete spectrum and continuum Weyl physics, 
as also discussed in Sec.~\ref{sec:weyl}.

To illustrate this behavior in higher angular momentum singular flat band model, we examine the case of $\ell=3/2$ with $\eta=100$, as shown in Fig.~\ref{fig:enhanced32}.
While Zeeman splitting is substantially enhanced, 
the spectrum in Fig.~\ref{fig:enhanced32} is 
related to that in Fig.~\ref{fig:higherl}(a) via the 
rescaling $E\rightarrow E/\eta$ and 
$k_z\rightarrow k_z/\sqrt{\eta}$. 
In this rescaled system, the Landau level spacing becomes relatively smaller compared to the band structure, making the underlying features more apparent.  
The two Weyl points along the $k_z$-axis in Fig.~\ref{fig:enhanced32}(a)
arises from a band inversion between states with angular momentum projections: one between $m_z=-1/2$ and $m'_z=-3/2$, 
and the other between $m_z=+1/2$ and $m'_z=-3/2$.
Owing to the correspondence between the Chern number and the total angular momentum projection $m_z$ along the high symmetry axis, 
the helicities of these Weyl points are given by $\chi=m_z-m_z'$, resulting in $\chi=1$ and $2$, respectively.

In the same manner as for $\ell=1$ in Fig.~\ref{fig:weyl}, 
$\ell=3/2$ system exhibits a single flat chiral Landau level associated with the $\chi=1$ Weyl point.  
This modes corresponds to $j=-3/2$ state, shown as the red line in Fig.~\ref{fig:enhanced32}(b).
Additionally, at the $\chi=2$ Weyl points located at larger $k_z$, the $j=-1/2$ mode (yellow curve in Fig.~\ref{fig:enhanced32}(b)) approaches and aligns with the $j=-3/2$ mode. 
Togather, they form two chiral Landau levels that connect states below and above the Weyl points.

This analysis can be systematically extended to arbitrary higher angular momentum $\ell$.
For general $\ell$, a series of Weyl points appears along $k_z$ axis, with helicities $\chi=1, 2, \cdots, 2\ell-1$ ordered increasingly from $k_z=0$ toward $k_z\rightarrow +\infty$.
For each $\ell$, a flat chiral Landau level emerges, associated with the $\chi=1$ Weyl points at the lowest $k_z$, and characterized by the total angular momentum $j=-\ell$ (See Fig.~\ref{fig:higherl}(d)).
As $k_z$ increases and passes through each subsequent Weyl point, 
a Landau level dispersion approaches and align with the flat landau levels. This results in multiple chiral Landau levels forming sequentially, each associated with a Weyl points of increasing helicity $\chi$ at larger $k_z$.

\bibliography{reference}
\end{document}